\documentclass[a4paper,11pt]{article}
\pdfoutput=1 % if your are submitting a pdflatex (i.e. if you have
\usepackage{jcappub} % for details on the use of the package, please
\usepackage[T1]{fontenc}
\usepackage{epsfig,latexsym,amssymb,amsmath}

\usepackage{hyperref}      % hypertext links %%ARXIV
\usepackage{graphicx}
\usepackage{times,psfrag,rotating}

\usepackage[caption=false]{subfig}
\usepackage{epstopdf}
\usepackage{color}
\usepackage{bm}
\usepackage[normalem]{ulem}
\usepackage{array}
\usepackage{multirow, multicol}

\renewcommand{\thetable}{$\arabic{table}$}

\newcolumntype{A}{ >{\centering\arraybackslash} m{3.5cm} }
\newcolumntype{B}{ >{\centering\arraybackslash} m{2.5cm} }
\newcolumntype{C}{ >{\centering\arraybackslash} m{2cm} }
\newcolumntype{D}{ >{\centering\arraybackslash} m{1.5cm} }
\newcolumntype{E}{ >{\centering\arraybackslash} m{2.5cm} }

\def\HI{H\,{\sc i}~}
\def\HII{H\,{\sc ii}~}

\def\xb{\bar{x}_{{\rm H\,\textsc{\scriptsize i}}}}

\def\beq{\begin{equation}}
\def\eeq{\end{equation}}
\def\bea{\begin{eqnarray}}
\def\eea{\end{eqnarray}}

\begin{document}
\hfill{\small IITB-PHY11122017} \\

\title{On dark matter - dark radiation interaction and cosmic reionization}

\affiliation[a]{Indian Institute of Astrophysics, Bengaluru -- 560034, India}
\affiliation[b]{Astronomy Centre, Department of Physics and Astronomy, University of Sussex, Brighton, BN1 9QH, U.K.}
\affiliation[c]{Department of Physics and Centre for Theoretical Studies, Indian Institute of Technology Kharagpur, Kharagpur -- 721302, India}
\affiliation[d]{Department of Physics, Indian Institute of Technology - Bombay, Powai, Mumbai 400076, India}

\author[a]{Subinoy Das,}
\emailAdd{subinoy@iiap.res.in}

\author[b,c]{Rajesh Mondal,}
\emailAdd{rm@phy.iitkgp.ernet.in}

\author[d]{Vikram Rentala,}
\emailAdd{rentala@phy.iitb.ac.in}

\author[a]{Srikanth Suresh}
\emailAdd{srikanthsrsh@gmail.com}

\abstract{
The dark matter sector of our universe can be much richer than the conventional picture of a single weakly interacting cold dark matter particle species. An intriguing possibility is that the dark matter particle interacts with a dark radiation component. If the non-gravitational interactions of the dark matter and dark radiation species with Standard Model particles are highly suppressed, then astrophysics and cosmology could be our only windows into probing the dynamics of such a dark sector. It is well known that such dark sectors would lead to suppression of small scale structure, which would be constrained by measurements of the Lyman-$\alpha$ forest. In this work we consider the cosmological signatures of such dark sectors on the reionization history of our universe. Working within the recently proposed ``ETHOS'' (effective theory of structure formation) framework, we show that if such a dark sector exists in our universe, the suppression of low mass dark matter halos would also reduce the total number of ionizing photons, thus affecting the reionization history of our universe. We place constraints on the interaction strengths within such dark sectors by using the measured value of the optical depth from the Planck satellite, as well as from demanding a successful reionization history. We compare and contrast such scenarios with warm dark matter scenarios which also suppress structure formation on small scales. In a model where dark matter interacts with a sterile neutrino, we find a bound on the ETHOS parameter $a_4 \lesssim 1.2\times 10^6 \textrm{ Mpc}^{-1}$. For warm dark matter models, we constrain the mass $m_{\textrm{WDM}} \gtrsim 0.7 \textrm{ keV}$, which is comparable to bounds obtained from Lyman-$\alpha$ measurements. Future 21-cm experiments will measure the global history of reionization and the neutral hydrogen power spectrum, which could either lead to stronger constraints or discovery of secret dark sector interactions. }

%\pacs{95.35.+d,96.50.S-,98.80.Cq,95.85.Ry}
\maketitle

{\let\clearpage\relax\tableofcontents}

%%%%%%%%%%%%%%%%%%%%%%%%%%%%%%%%%%%%%%%%%%%%%%%%%%%%%%%%%%%%%%%%%%%%%%%%
\section{Introduction}
In spite of intensive search over decades, the nature of dark matter (DM) still remains a mystery. The simple hypothesis of a cold dark matter (CDM) particle that interacts extremely weakly with the Standard Model (SM) particles is in good agreement with astrophysical observations on many scales ranging from galaxies \cite{Rubin:1970zza}, to cluster of galaxies \cite{Randall:2007ph}, to cosmological observations of large scale structure \cite{Tegmark:2003ud} and the cosmic microwave background (CMB)~\cite{Ade:2015xua}. In fact all of our evidence for the existence of DM comes from astrophysical observations of its \textit{gravitational} interactions. However, if dark matter really does behave like other particles that we are familiar with, we would expect the DM particle to have other non-gravitational interactions. Thus the fundamental objective of DM particle physics research is the search for such non-gravitational interactions.

Traditional candidates for the DM particle have focussed on particles with the right properties to behave as DM, which also solve certain theoretical or experimental problems with the SM. For example, Weakly Interacting Massive Particles (WIMPs) appear in many extensions of the SM which attempt to address the hierarchy problem~\cite{Bertone:2004pz, Feng:2010gw}. The typical strategies for detecting such WIMPs rely on looking for non-gravitational interactions of these particles with SM particles~\cite{Jungman:1995df}. For example, direct detection searches focus on looking for WIMP DM particles passing through the earth and interacting with nuclei in a detector~\cite{Goodman:1984dc, Undagoitia:2015gya}. Currently, there is no convincing evidence for the existence of such WIMPs through the probes that we have designed, but this allows us to place strong upper limits on the existence of interactions between DM and SM particles. As we approach the limit of the ``neutrino floor'' with the next generation of DM direct detection experiments (where coherent neutrino scattering starts to become a large background in the detector)~\cite{Monroe:2007xp, Gutlein:2010tq}, we may be approaching a natural limit in our ability to search for WIMP dark matter. While the particle physics community could take the approach of focussing other DM candidates such as axions~\cite{Duffy:2009ig}, sterile neutrinos~\cite{Merle:2017jfn}, fuzzy dark matter~\cite{Hu:2000ke}, lighter WIMPS~\cite{Alexander:2016aln} etc., it is also a time for introspection for the paradigm of terrestrial DM searches as a whole.

It may well be the case that DM interactions with the SM are highly suppressed and attempts to directly search for DM in terrestrial experiments may result in a dead-end. In such a scenario, one may ask if there is any way to bolster the evidence for a particle DM species that permeates the universe. One intriguing possibility is that the dark sector could be much richer than conventionally imagined. A simple scenario is a dark sector with a DM particle coupled to a light (nearly massless) species that plays a role analogous to that of photons/neutrinos in the visible (SM) sector. Such a species has been dubbed ``dark radiation'' (DR) and many scenarios for such a dark sector have been proposed in the particle physics literature~\cite{Ackerman:mha, Feng:2009mn, Tulin:2017ara}. From a UV perspective, such a dark sector is motivated by the ``hidden valley'' scenarios~\cite{Strassler:2006im, ArkaniHamed:2008qn}.

The existence of dark radiation opens up several probes for searches for dark sector dynamics. Since the direct interactions of such a dark sector with the SM particles are assumed to be very weak, the dominant signatures of such a dark sector would be astrophysical and cosmological. An extra radiation degree of freedom is usually parameterized by a number of effective neutrino species and is constrained by observations of the CMB and the well established theory of big bang nucleosynthesis (BBN). However, if the radiation degree of freedom is \textit{colder} by a factor of a few compared to the cosmic neutrino background formed out of active neutrino species, then such constraints would be easily evaded \cite{Dasgupta:2013zpn}. An interesting observational consequence of a rich dark sector is the impact on galactic halos due to dark matter self-interactions mediated via the dark radiation species.

There are several discrepancies that arise on small scales when making comparisons between simulated, collisionless DM halos and the observationally inferred properties of these halos. Some of the well known problems are the missing satellites problem~\cite{Klypin:1999uc, Moore:1999nt, Foot:2016wvj}, the too-big-to-fail problem~\cite{BoylanKolchin:2011de,2012MNRAS.423.3740V} and the core-cusp problem~\cite{Oh:2010ea, Rocha:2012jg, Peter:2012jh, 2013MNRAS.431L..20Z, Foot:2014uba}. Several astrophysical solutions have been proposed as resolutions to these problems. For example the ``missing satellite problem'' could be explained by our failure to detect ultra-faint dwarf satellites \cite{Tollerud:2008ze} of the Milky Way. Other studies have pointed out that simple $N$-body simulations used to study DM on small scales are inadequate since they do not account for effects of baryonic physics which could significantly alter the distribution of DM \cite{Bullock:2000wn,Benson:2001at,2012MNRAS.422.1231G}. For a recent review on the status of these problems and potential astrophysical resolutions see~\cite{DelPopolo:2016emo}.

While astrophysical explanations could be the resolution to these small scale structure problems, there has also been extensive work on DM self-interactions as a possible solution to these problems (for a recent review of such explanations see~\cite{Tulin:2017ara}). Self-interacting DM scenarios are subject to constraints from merging clusters such as the bullet cluster~\cite{Randall:2007ph}, as well as the observed ellipticity of DM halos on cluster scales as measured through strong gravitational lensing~\cite{MiraldaEscude:2000qt, Meneghetti:2000gm}, although the latter constraints have been argued to be less robust~\cite{Rocha:2012jg,Peter:2012jh}. Reconciling the large self-interaction cross-sections required to explain the small scale structure problems with the bounds from merging clusters typically leads to the conclusion that the DM self-interaction cross-section  must have an inverse velocity dependence. This kind of scattering can occur if DM is coupled to a light or massless mediator~\cite{Feng:2009mn, Agrawal:2016quu}. Such a mediator would play the role of DR in the early universe.

On cosmological scales the DM-DR interactions would have an impact on structure formation. A dark radiation component would create a source of pressure in the coupled DM-DR fluid in the early universe and this would prevent DM over-densities from collapsing efficiently. This behavior is familiar from the behavior of the baryon-photon fluid where acoustic waves can be set up in the coupled fluid. Similarly, a complex dark sector would also experience Dark Acoustic Oscillations (DAO) along with collisional or silk-damping effects in the DM-DR fluid~\cite{Loeb:2005pm, Cyr-Racine:2013fsa, Buckley:2014hja}. Even though the DR would carry an imprint of these acoustic oscillations, the cosmic DR background (the analogue of the CMB in the dark sector) is likely to be unobservable. However the imprint of the acoustic oscillations and damping would also show up in the linear matter power spectrum of dark matter, which would in turn have an effect on the visible matter distribution. While non-linearities would smooth out the power somewhat, some relics of such an imprint (especially the suppression of power on small scales) would remain. Such a scenario has been constrained from observations of structure on small scales using Lyman-$\alpha$ data~\cite{Murgia:2017lwo}. While we expect qualitative similarities to warm dark matter (WDM) scenarios where structure formation on small scales is suppressed due to the free-streaming of dark matter~\cite{Colin:2000dn}, there are some important differences due to the fact that acoustic oscillations are seen in the power spectrum rather than just a smooth damping relative to the power spectrum of standard $\Lambda$CDM scenarios.

In the present work, we will examine the impact of DAO on cosmic reionization. With suppressed power on small scales, this would reduce the count of smaller halos in the early universe. The star light from galaxies in these halos is expected to provide ionizing UV radiation which reionizes the universe around a redshift $z \sim 10$. Thus, any suppression of small scale structure would delay or change the reionization history of the universe. By demanding consistency with optical depth measurements from Planck~\cite{Ade:2015xua} and the observation that the universe is completely reionized by a redshift of $z=6$ \cite{2006AJ....132..117F}, we will demonstrate that one can obtain a bound on the DM-DR interaction strength\footnote{For similar attempts at constraining WDM scenarios using cosmic reionization, see Refs.~\cite{Tan:2016xvl, Villanueva-Domingo:2017lae, Lopez-Honorez:2017csg}.}. Ongoing and future 21-cm surveys of the neutral Hydrogen distribution at high redshifts ($z > 8$) will also probe the global history of reionization and the 21-cm power spectrum~\cite{2013ExA....36..235M,Pober:2013jna, Mesinger:2015sha} and we will speculate on future constraints/imprints that could be seen in such surveys.

This paper is organized as follows: In Sec.~\ref{sec:ETHOSreview}, we review the basic ETHOS~\cite{Cyr-Racine:2015ihg, Vogelsberger:2015gpr} framework for studying the effect on the matter power spectrum in dark matter sectors with a dark radiation component. In Sec.~\ref{sec:qualitative} we will give both qualitative as well as quantitative (numerical/approximate analytic) results for the impact of such dark sectors on the suppression of the linear matter power spectrum on small scales and highlight the differences with WDM scenarios. A more detailed explanation for the analytic formulae presented in this section is provided in Appendix~\ref{sec:appendix}. In Sec.~\ref{sec:result} we will examine the impact of the suppression of small scale structure on the reionization of the universe.
In Sec.~\ref{sec:resulta}, we show that the suppression on small scales persists even when taking non-linearities into account, then in Sec.~\ref{sec:resultb}, we generate neutral hydrogen maps of the universe around the epoch of reionization and set bounds on the range of ETHOS and WDM model parameters by demanding consistency with a successful reionization history.
We discuss future constraints and observational prospects of DM-DR interactions using 21-cm measurements in Sec.~\ref{sec:21cm}.
Finally, we conclude and discuss future directions in Sec.~\ref{sec:conclusions}.

%%%%%%%%%%%%%%%%%%%%%%%%%%%%%%%%%%%%%%%%%%%%%%%%%%%%%%%%%%%%%%%%%%%%%%%%%%%%%%
\section{Review of the ETHOS Framework}
\label{sec:ETHOSreview}
We begin by briefly reviewing the framework of ETHOS~\cite{Cyr-Racine:2015ihg, Vogelsberger:2015gpr} which greatly simplifies the analysis of cosmological perturbations evolved in the presence of a non-minimal dark sector. In the ETHOS framework, one can map the parameters in the Lagrangian of a dark sector with interacting dark matter and dark radiation into a series of coefficients that appear in a redshift series expansion of the collision terms of the Boltzmann equations for DM and DR. Solving the equations in this framework allows one to easily map the impact of dark sector self-interactions to the temperature and density fluctuations in the early universe and in particular to derive a modified matter power spectrum. We note that the ETHOS framework is general enough to allow for interactions between DM and DR as well as DR self-scattering. For this work we will only consider DM-DR scattering and assume that DR self-scattering is negligible or absent.

The particle physics parameters (couplings and masses) enter the Boltzmann equation describing the phase space evolution of a species (DM or DR) through ``opacity coefficients'' ($\dot{\kappa}$) that are related to the mean free path ($\lambda$) of the relevant particle by,
\begin{equation}
-\dot{\kappa} \simeq 1/\lambda \simeq (n \sigma).
\end{equation}
Here, $n$ is the number density of scattering targets and $\sigma$ is the (thermal) cross-section of the relevant species with the target. For the type of interactions we are considering, we have two opacity coefficients ($\dot{\kappa}_{\chi}$ and $\dot{\kappa}_{\textrm{DR-DM}}$)  describing respectively the opacity for DM scattering off of DR and DR scattering off of DM.

In the ETHOS parametrization the scaling of $\dot{\kappa}_{\chi}$ with redshift $(z)$ is given in a power series expansion by,
\begin{equation}
\dot{\kappa}_{\chi} = - \frac{4}{3} \Omega_{\textrm{DR}} h^2 \sum_n  a_n \frac{(1+ z)^{n+1}}{(1+z_D)^n}.
\end{equation}
Here, $a_n$ are model dependent coefficients (with dimensions of length$^{-1}$) that depend on the DM-DR scattering amplitude, $\Omega_{\textrm{DR}}$ is the present-day dark radiation relic density fraction and $h$ is the usual dimensionless Hubble parameter. Here, $z_D$ is a fixed parameter degenerate with the normalization of the $a_n$. We choose it to be $10^7$ for numerical convenience. The redshift power-series expansion of $\dot{\kappa}_{\textrm{DR-DM}}$ is given in terms of the same coefficients $a_n$ multiplied by powers of $(1+ z)^n$ and the cosmological DM density fraction $\Omega_{\textrm{DM}} h^2 = 0.12206$,
\begin{equation}
\dot{\kappa}_{DR-DM} = - \Omega_{\textrm{DM}} h^2 \sum_n  a_n \frac{(1+ z)^n}{(1+z_D)^n}.
\end{equation}

In order to describe the evolution of the density perturbations one needs to also fix the value of the present day DR energy density fraction $\Omega_{\textrm{DR}}$ (or equivalently the present day temperature ratio of the DR and CMB, $\xi \equiv T_{\textrm{DR}}/T_{\textrm{CMB}} |_{z=0}$\footnote{As long as we are studying the density perturbations after $e^\pm$ pairs have transferred their energy to the photon bath (which occurs at a redshift of $z \sim 10^9$), and if we additionally assume that no new degrees of freedom in the dark sector annihilate and dump their energy into the DR plasma, the temperature ratio of these two sectors will not vary with redshift.}).

Thus the coupling constants and masses of the particle physics model are mapped on to a series of coefficients $a_n$ (for $n \geq 0$), which are red-shift series expansion coefficients of the various opacities. We note that the DM heating rate, also has a redshift series expansion with coefficients denoted as $d_n$ (see Eq. 15 in Ref.~\cite{Cyr-Racine:2015ihg}). The higher moments of the DR perturbations (quadrupole and above) are also coupled to each other (and to the density and velocity perturbations) through angle dependent scattering of DR with DM. The evolution equation for the $l$-th moment ($l \geq 2$) has coefficients $\alpha_l$ multiplying the DR-DM opacity. The coefficients $\alpha_l$ are determined from the angular dependence of the relevant particle physics amplitudes (see Eq.~3 and Eq.~6 in Ref.~\cite{Cyr-Racine:2015ihg}).

Thus in summary, an ETHOS model with only DM-DR interactions is a mapping from the particle physics model of the dark sector to the set of parameters $\{ \xi, a_n, d_n, \alpha_l \}$ which determine the evolution of cosmological perturbations. In practice, the coefficients $d_n$ and $\alpha_l$ for $l > 2$ have very little effect on the matter power spectrum, so we will not explicitly state these parameters. The main advantage of the ETHOS framework is that for typical particle physics models, only a few of the ETHOS coefficients are non-zero, thus greatly simplifying cosmological calculations.

\subsection{ETHOS Models}
\label{sec:ethosmodels}
We will consider a specific model of DM which has interactions with a DR component from the toy models specified in the ETHOS paper (we will refer to this model as ETHOS 1 or ETHOS model 1). We will also reproduce below the explicit mapping between the Lagrangian parameters and the ETHOS parameters.

\subsubsection{ETHOS 1: DM scattering with a sterile neutrino through a massive mediator}
In this model the DM species is a Dirac fermion $\chi$, which interacts with a (nearly massless) sterile neutrino $\nu_s$ (which plays the role of DR) through a heavy intermediary massive vector boson $\phi_\mu$\footnote{A similar model was studied in Ref.~\cite{Binder:2016pnr} where the authors examined the impact of DM-DR interactions on the linear matter power spectrum.}. The interaction Lagrangian is given by:
\begin{equation}
\mathcal{L} _{\textrm{int}} = -\frac{1}{2} m^2_\phi \phi_\mu \phi^\mu - \frac{1}{2} m_{\chi} \overline{\chi} \chi -g_\chi \phi_\mu \overline{\chi} \gamma^\mu \chi - \frac{1}{2} g_\nu \phi_\mu \overline{\nu}_s \gamma^\mu \nu_s.
\end{equation}

In this model the non-zero ETHOS parameters defined in terms of the Lagrangian parameters and the temperature ratio $\xi$ are:
\begin{align}
\Omega_{\textrm{DR}}h^2 &= 1.35 \times 10^{-6}  \left( \frac{\xi}{0.5} \right )^4, \nonumber \\
a_4 &= (1 + z_D)^4 \frac{3\pi}{2}\frac{ g^2_\chi g^2_\nu}{m^4_\phi} \frac{\rho_c/h^2}{m_\chi} \left( \frac{310}{441} \right) \xi^2 T^2_{CMB,0}, \nonumber  \\
&= 0.6 \times 10^5  \left(\frac{ g_\chi}{1} \right )^2 \left(\frac{ g_\nu}{1} \right )^2 \left( \frac{0.468 \textrm{ MeV}}{m_\phi} \right )^4 \left ( \frac{2 \textrm{ TeV}}{m_\chi} \right ) \left(\frac{ \xi}{0.5} \right )^2 \textrm{ Mpc}^{-1} , \nonumber  \\
\alpha_{l \geq 2} &= \frac{3}{2}.
\end{align}

In the above formulae, the expression for $a_4$ is valid as long as the DR temperature is smaller than the mediator mass ($m_\phi$). For the small benchmark value of the mediator mass chosen above, we note that both the mediator and the sterile neutrino dark radiation would contribute to the number of light degrees of freedom at BBN. However, for our choice of benchmark temperature ratio $\xi = 0.5$, this would still be safe from the BBN bounds on extra radiation species~\cite{Binder:2016pnr}. Once the temperature of the DR bath drops below the mass of the mediator, only the sterile neutrino would behave like a dark radiation species.

We note here that in this model and the other toy models specified in the ETHOS paper, typically only one of the $a_n$ coefficients is non-zero. The effect of DM-DR interactions on the matter power spectrum is most sensitive to this parameter. In the rest of this paper, we will assume a fixed value of $\xi = 0.5$ and we will focus on the dependence of the matter power spectrum on the $a_n$ coefficients.

\section{The linear matter power spectrum in ETHOS models}
\label{sec:qualitative}

\subsection{Qualitative understanding of evolution of linear density perturbations in ETHOS models}
We will first attempt to understand the impact of ETHOS parameters on the linear matter power spectrum. The key criteria is to determine the evolution of the Jeans scale in different cosmological epochs~\cite{2010gfe..book.....M}. Density perturbations on scales smaller than the Jeans scale are prevented from collapsing and thus the linear matter power spectrum on such small scales would be suppressed. At early times while DM is coupled to DR, dark acoustic oscillations give rise to oscillatory features in the power spectrum with peaks on scales with wave numbers,
\begin{equation}
k_p =  \frac{m \pi}{r^\textrm{max}_s}, \textrm{ for $m$ = 1, 3, 5 etc.}
\end{equation}
where $r^\textrm{max}_s$ is the sound horizon scale at DR decoupling which is given by,
\begin{equation}
\label{eq:maxSH1}
r^\textrm{max}_s  \simeq 0.18  \left ( \frac{a_4}{0.6 \times 10^{5} \textrm{ Mpc}^{-1}} \right )^{1/3} \textrm{ Mpc}.
\end{equation}
Once DR decouples from DM it free streams and continues to drag the DM until the DM decouples. This leads to a damping of the linear power spectrum on scales set by the DM drag scale which we can (over) estimate as the maximum free streaming scale of the DR,
\begin{equation}
\label{eq:maxjeans1}
\lambda_\textrm{fs}  \simeq 0.39 \left ( \frac{a_4}{0.6 \times 10^{5} \textrm{ Mpc}^{-1}} \right )^{1/4} \textrm{ Mpc}.
\end{equation}

The suppression of the power spectrum on small scales will lead to a suppression of low mass halos with masses smaller than the Jeans mass ($M_J$) which is given by,
\begin{equation}
M_J \simeq 10^{10} M_\odot \left ( \frac{a_4}{0.6 \times 10^{5} \textrm{ Mpc}^{-1}} \right )^{3/4},
\end{equation}
where $M_\odot$ denotes solar mass.

It is interesting to compare this to WDM models that also predict a suppression of the matter power spectrum on small scales. In WDM models the suppression scale is set by the maximum free streaming scale of WDM which is determined by the onset of non-relativistic free streaming,
\begin{equation}
\label{eq:WDMfsscale1}
\lambda_\textrm{fs}^{\textrm{WDM}} \simeq 0.15 \left(\frac{2 \textrm{ keV}}{m_\textrm{WDM}} \right) \textrm{ Mpc},
\end{equation}
where $m_\textrm{WDM}$ is the mass of the WDM candidate.

A detailed explanation of the scaling relations for $r^\textrm{max}_s$ and $\lambda_\textrm{fs}$ with the ETHOS parameter $a_4$ and for $\lambda_\textrm{fs}^{\textrm{WDM}}$ with $m_\textrm{WDM}$ is given in Appendix~\ref{sec:appendix}.

\subsection{Comparison with the numerical linear power spectrum}

\begin{figure}[h]
\centering
\includegraphics[width=1.05\textwidth]{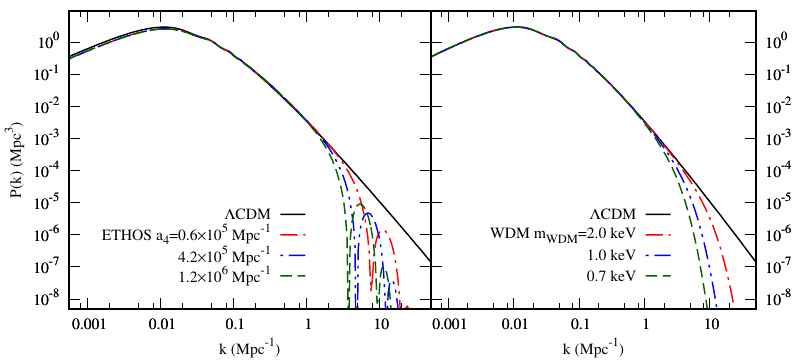}
\caption{Normalized linear matter power spectrum $P(k)$ as a function of $k$ at $z = 124$ for ETHOS models with $a_4 =\{0.6\times 10^5,\ 4.2\times 10^5,\ 1.2\times 10^6 \} \textrm{ Mpc}^{-1}$ and WDM models with $m_{\textrm{WDM}}=\{2.0,\ 1.0,\ 0.7\}\,{\rm keV}$. The suppression of the power spectrum on small scales relative to the standard $\Lambda$CDM case is clearly visible. Dark acoustic oscillation (DAO) features are also visible in the ETHOS models that are absent in the WDM models.}
\label{fig:1}
\end{figure}

The qualitative/analytic predictions for the matter power spectrum can be checked by numerically computing the linear power spectrum in both the ETHOS and WDM scenarios. We consider three benchmark model points for ETHOS with $a_4 =\{0.6\times 10^5,\  4.2\times 10^5,\ 1.2\times 10^6\} \textrm{ Mpc}^{-1}$ and three benchmark WDM models with $m_{\textrm{WDM}}=\{2.0,\ 1.0,\ 0.7\}\,{\rm keV}$. We fix $\xi=0.5$ for all three ETHOS model points. To compute the numerical power spectra for the ETHOS models, we use the publicly available ETHOS-CAMB code~\cite{Lewis:1999bs,Cyr-Racine:2015ihg} and for the WDM models, we have used the analytical formula in Ref.~\cite{Bode:2000gq}. The results are shown in Fig.~\ref{fig:1}.

The left panel of Fig.~\ref{fig:1} shows the linear matter power spectrum at $z=124$ for the three ETHOS models and the standard $\Lambda$CDM power spectrum. We can clearly see the suppression of power on small scales along with the oscillation peaks. The right panel shows the same power spectrum for the WDM models. In this case, we see a smooth suppression of power on small scales without oscillation peaks. The relevant damping and oscillation scales are well predicted by our analytic formulae. Eq.~\ref{eq:maxSH1} and Eq.~\ref{eq:maxjeans1} predict the relevant oscillation and damping scales respectively in the ETHOS models and Eq.~\ref{eq:WDMfsscale1} predicts the damping scale for WDM models. We see that as the value of $a_4$ is increased (more scattering of DM and DR) in the ETHOS models or when $m_{\textrm{WDM}}$ is decreased (later onset of non-relativistic free-streaming) in WDM models, we get suppression on larger length scales (smaller values of $k$), relative to the standard $\Lambda$CDM scenario.

\section{Impact of suppression of small scale power spectrum on reionization}
\label{sec:result}
The suppression of the matter power spectrum on small scales could have a dramatic impact on the reionization history of the universe. Our goal is to study the observable impact of the suppression of the small scale power spectra in ETHOS and WDM models during the Epoch of Reionization (EoR) and to place constraints on the model parameters from current observations.

Schematically, the connection between suppression of small scale structure and reionization is as follows: First, we note that non-linearities which govern the late time evolution of the cosmic density perturbations will smooth out the power spectrum to some extent. However, we still expect a suppression in the power spectrum on small length scales (large $k$) relative to the standard $\Lambda$CDM picture. This in turn leads to suppressed formation of lower mass halos. The deficit of low mass halos leads to reduced reionization of the universe due to reduced galaxy formation.

There are two main observational constraints that must be satisfied in such scenarios. The first constraint arises from the measurement of the optical depth to reionization $(\tau_{\textrm{reio}})$, which can be inferred from CMB anisotropy and polarization measurements. These CMB constraints can be satisfied for a wide range of ionization histories~\cite{Robertson:2015uda}. It is instructive to note that from the latest Planck optical depth measurements $(\tau_{\textrm{reio}} = 0.058 \pm 0.012)$~\cite{Adam:2016hgk}, early reionization (before $z \sim 10$) is disfavored. The second observation from quasar absorption spectra is the Gunn-Peterson bound \cite{becker01,fan03,becker15}, which tells us that reionization is almost complete by $z=6$.

The global evolution of the mass-averaged neutral hydrogen (\HI) fraction $\xb(z)$ as a function of redshift during the EoR is largely unconstrained from either of the constraints referred to above. We can therefore take a simplified version of these two constraints, by requiring that $\xb(z=8)=0.5$, i.e. reionization is $50\%$ complete by $z=8$. In practice, for most scenarios, this simplified criteria suffices to satisfy both the optical depth constraints and the Gunn-Peterson bound.

In this section, we will derive a constraint on the ETHOS and WDM model parameter space by demanding consistency with the constraint  $\xb(z=8)=0.5$. In the next section we will also comment on future 21-cm observations of the neutral hydrogen which can probe the global history of reionization, as well as measure the power spectra of \HI, and could potentially place stronger constraints on such dark sectors.

\subsection{The non-linear power spectrum and the suppression of low mass halos}
\label{sec:resulta}
In order to assess the impact of suppression of small scale structure on reionization, we must first compute the non-linear matter power spectrum at $z=8$. We start with the linear power spectra (Fig.~\ref{fig:1}) for the three ETHOS and three WDM benchmark models and for the standard $\Lambda$CDM model at very high redshift around $z=124$. For each model we generate an initial Gaussian random density field at $z=124$. We then evolve the power spectra non-linearly till the epoch of reionization with our own $N$-body simulation code.

\begin{figure}[h]
\centering
\includegraphics[width=1.05\textwidth]{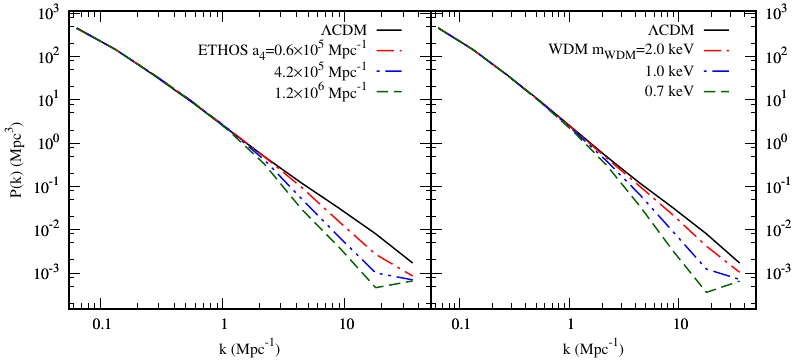}
\caption{The figures above show the non-linear matter power spectrum $P(k)$ as a function of $k$ at $z=8$ calculated from the result of an $N$-body simulation using an initial Gaussian random density field based on the linear power spectrum for each of the three different benchmark ETHOS and WDM models. We can see that the suppression of small scale power persists in the non-linear power spectrum, although non-linearities smooth out the oscillatory features of the ETHOS power spectrum. The black solid curves show the $\Lambda$CDM power spectrum.}
\label{fig:pk}
\end{figure}

We use a particle mesh $N$-body code to generate the dark matter distribution at $z=8$.  We have run simulations with comoving volume $[150\, {\rm Mpc}]^3$ with a $2144^3$ grid using $1072^3$ dark matter particles. The spatial resolution of our grid is $0.07\, {\rm Mpc}$ which corresponds to a mass resolution of $1.09 \times10^8\, {\rm M_{\odot}}$. Fig.~\ref{fig:pk} shows the matter power spectrum $P (k)$ output from the $N$-body simulation at $z=8$. Note the suppression of power on scales $k \gtrsim 1\,{\rm Mpc}^{-1}$ for the ETHOS (left panel) and WDM (right panel) models compared to the $\Lambda$CDM model. This suppression deepens as we increase the value of $a_4$ in ETHOS model 1 or with the decrease of $m_{\textrm{WDM}}$ in the WDM model. Also, we see that the non-linearities smooth out the oscillation peaks in the linear power spectrum of the ETHOS models.

In the next step, we use the Friends-of-Friends (FoF) algorithm to identify collapsed halos in the dark matter distribution. We have used a fixed linking length of $0.2$ times the mean inter-particle distance and also set the criterion that a halo should have at least $10$ dark matter particles, which implies a minimum halo mass of $1.09 \times10^9\, {\rm M_{\odot}}$. Our choice of the minimum halo mass is well motivated because a halo mass of a few $10^8\,{\rm M_{\odot}}$ \cite{Mesinger:2007pd} (at $z=8$) also corresponds to the virial temperature ($\sim 10^4$~K) of the \HI-cooling threshold for star formation.

\begin{figure}[h]
\centering
\includegraphics[width=1.05\textwidth]{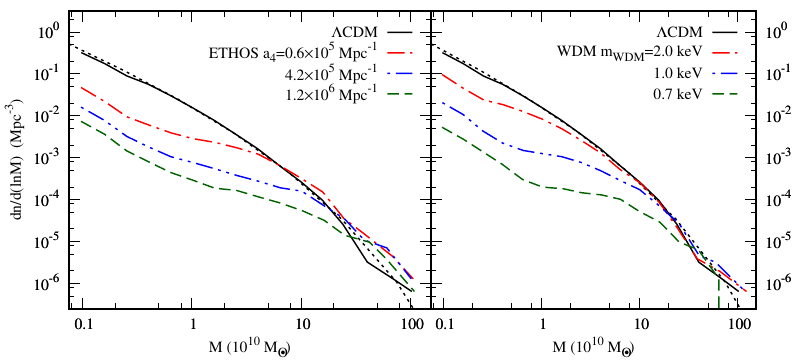}
\caption{The figures above show the halo mass function extracted from our $N$-body simulations using the Friends-of-Friends (FoF) algorithm for each of the three benchmark ETHOS and WDM models. As expected low mass halos are suppressed in the ETHOS and WDM models as compared to the $\Lambda$CDM (solid black lines) scenario. The black dotted curve is the theoretical $\Lambda$CDM mass function of Ref.~\cite{Sheth:2001dp} which is in good agreement with our simulated $\Lambda$CDM mass function.}
\label{fig:mass_func}
\end{figure}

Fig.~\ref{fig:mass_func} shows the simulated comoving number density of halos per unit logarithmic halo mass ${\rm d}n/{\rm d(ln}M)$ as a function of the halo mass $M$ at $z=8$ for the benchmark ETHOS and WDM model parameters that we consider here. The black dashed curves show the theoretical $\Lambda$CDM mass function assuming a spherical collapse model~\cite{Sheth:2001dp}. The simulated $\Lambda$CDM mass function, shown with solid black lines, is in very good agreement with the theoretical mass function. Note that the low mass halo abundance is substantially reduced for the ETHOS and WDM models as compared to the $\Lambda$CDM model. Since, the linear density perturbations on small scales in the early universe seed the formation of low mass halos, the number counts for low mass halos decreases as the linear matter power spectrum on small scales is reduced.

\subsection{Impact on reionization}
\label{sec:resultb}
\subsubsection{The parameter $N_{\textrm{ion}}$}
To assess the impact of the halo abundance on the ionization of the universe at a particular redshift, one needs to assume a relationship between the halo mass ($M_\textrm{halo}$) and the number of ionizing photons emitted per halo ($N_\gamma^{\textrm{halo}}$). Typically, one assumes a linear relationship of the form \cite{Choudhury:2008aw},
\begin{equation}
N_\gamma^{\textrm{halo}} = N_{\textrm{ion}} \frac{M_\textrm{halo}}{m_H}.
\label{eq:Nion1}
\end{equation}
Here, $N_{\textrm{ion}}$ is a proportionality factor that gives us the number of ionizing photons emitted per unit halo mass, where we measure the halo mass in units of a reference baryon (hydrogen) mass $m_H$. The parameter $N_{\textrm{ion}}$ will be crucial for generating the connection between halos and the reionization signal in our simulations.

We can estimate $N_{\textrm{ion}}$ roughly as follows from the following relationship \cite{2016MNRAS.459.2342M},
\begin{equation}
 N_\textrm{ion} = 8
 \left( \frac{N^\textrm{b}_\textrm{ion}}{4000} \right)
 \left( \frac{M_\textrm{b}/M_\textrm{halo}}{1/5} \right)
  \left( \frac{\epsilon_\textrm{esc}}{10\%}  \right )
\left ( \frac{\epsilon_\textrm{SF}}{10\%} \right ).
\end{equation}
Here, $N^\textrm{b}_\textrm{ion}$ is the number of ionizing photons produced in the halo per baryon (assumed to be all hydrogen). This factor can be estimated for baryons which are processed inside stars. $M_\textrm{b}/M_\textrm{halo}$ is the baryonic mass fraction. The total number of ionizing photons that escape the galaxy is tempered by an escape fraction $\epsilon_\textrm{esc}$ that characterizes the transparency of the halo to ionizing radiation. In addition, not all baryons will take part in star formation, so a proportionality factor for the star formation  efficiency (denoted by $\epsilon_\textrm{SF}$) must also be included. In the relationship above, we have assumed some nominal values for normalization of the above parameters, each of which is subject to large systematic uncertainties.

Now the question is: What are acceptable values of $N_{\textrm{ion}}$ if ionizing photons from star-forming galaxies are responsible for the reionization process? Just to give an estimate, it is known that for a metallicity $Z = 0.01$ and Scalo stellar mass function \cite{1986FCPh...11....1S}, $N^\textrm{b}_\textrm{ion}$ is nearly 4000 per baryon. It should be noted that all these factors: metallicity, initial mass function, star formation efficiency, and escape fraction, are highly uncertain\footnote{For a recent take on the predictions and uncertainties for these quantities, and how the James Webb Space Telescope is expected to improve upon these, see for example Ref.~\cite{doi:10.1093/mnras/stw980}.}~\cite{2000ApJ...542..548R,2000ApJ...545...86W,2008ApJ...672..765G,2009ApJ...693..984W,2001ApJ...546..665S,2006ApJ...651..688S,2006MNRAS.371L...1I,2007ApJ...667L.125C}.
The only way to obtain a high value of $N_{\textrm{ion}}$ is for zero metallicity (population III) stars where the number of photons per stellar baryon could be significantly higher and lie in the range $10^4$ to $10^5$, thus corresponding to higher $N_{\textrm{ion}}\sim\mathcal{O}(20-200)$. But these stars last only a few million years which is considerably smaller than the age of the universe and can not reionize the universe. In general we can safely say that $N_{\textrm{ion}} \lesssim 500$~\cite{Barkana:2000fd, 2009CSci...97..841C, 2016MNRAS.459.2342M}.

We note the key assumptions here in the definition of $N_{\textrm{ion}}$. The first assumption is the linear relationship between halo mass and $N_\gamma^{\textrm{halo}}$, this can be equivalently stated as the assumption that $N_{\textrm{ion}}$ is independent of the halo mass. The second is the assumption of redshift independence of $N_{\textrm{ion}}$.

\subsubsection{Generating the ionized and neutral hydrogen distribution}
Using different values for $N_{\textrm{ion}}$, we can now numerically find the impact on reionization from halos with a particular mass distribution. We will attempt to satisfy the constraint of achieving $50\%$ reionization at $z=8$.
Our expectation is straightforward. With a reduced abundance of low mass halos, we need larger values of $N_{\textrm{ion}}$ to generate similar levels of reionization. If the suppression of low mass halos is too drastic, then we will need unrealistic values of $N_{\textrm{ion}}$, thus ruling out such models. We describe below in detail the procedure used to generate 3-D ionization maps at a given redshift which is based on the excursion-set formalism of Ref.~\cite{Furlanetto:2004nh}.

 We first identify a field for the neutral hydrogen distribution, which we assume traces the matter distribution of our $N$-body simulations with a spatially homogenous fraction determined by the cosmological baryon abundance.
We then generate a halo density field on our grid from the halo distribution identified in our simulation by using the cloud-in-a-cell algorithm. We then generate an ionizing photon field, which is initially taken to coincide with the locations of the collapsed halos found in our simulation and whose number densities are obtained using Eq.~\ref{eq:Nion1}, for a given value of the parameter $N_{\textrm{ion}}$. Thus, we have a grid of the \HI distribution (which traces the matter) and a seed photon distribution (which traces the halos).

We then account for the propagation of the ionizing photons from the halos into the intergalactic medium (IGM) and the consequent ionization of the neutral hydrogen field. The propagation length of the photons is given by the mean free path $R_{\rm mfp}$, which we have set to 20~Mpc. This is consistent with the findings of Ref.~\cite{2010ApJ...721.1448S} from the study of Lyman limit systems at low redshifts. Our procedure for simulating the photon propagation is to iteratively smooth over the \HI field and the photon field using spheres of radius $R$ starting from the grid length to $R_{\rm mfp}$ using a top-hat filter. At each step, we determine whether a grid point is to be ionized if,
\begin{equation}
\langle n_\gamma(x) \rangle > \langle n_H(x) \rangle,
\label{eq:ioncriteria}
\end{equation}
where $n_\gamma(x)$ is the photon number density and $n_H(x)$ is the hydrogen number density at that grid point. Here, we have assumed that the effect of hydrogen recombination in the IGM is negligible although such effects could easily be taken into account by a trivial, position-independent rescaling of the right-hand-side of Eq.~\ref{eq:ioncriteria} in the so called ``homogenous recombination scheme'' (see Ref.~\cite{Choudhury:2008aw}). We use this ionization criteria to define an \HII field in our simulation box. After all iterations of the photon propagation are completed, we are thus left with a neutral as well as ionized hydrogen map. Adding up the total amount of \HII and comparing to the total hydrogen gives us the global ionization fraction at a given redshift. The grid used to generate the ionization maps are eight times coarser than the one used for the $N$-body simulations. The steps outlined in this paragraph closely follow Refs.~\cite{2014MNRAS.443.2843M,2015MNRAS.449L..41M,2016MNRAS.456.1936M,2017MNRAS.464.2992M,2018MNRAS.474.1390M}.

It is possible to achieve different reionization histories $\xb(z)$ by varying the parameter $N_{\textrm{ion}}$. For each benchmark model, we adjust the value of $N_{\textrm{ion}}$ to achieve $50\%$ reionization at $z=8$ in order to satisfy the observational constraints. The value of $N_{\textrm{ion}}$ needed is $23$ for the $\Lambda$CDM model. For the ETHOS models, as $a_4$ increases, larger scales undergo suppression (Eq.~\ref{eq:maxjeans1}) and therefore we need correspondingly larger values of $N_{\textrm{ion}}$ in order to achieve the same level of ionization at $z=8$. The values (corresponding to increasing values of the benchmark $a_4$ points) are given by $N_{\textrm{ion}} = \{100,\ 301,\ 721\}$. Similarly, in the WDM models as $m_{\textrm{WDM}}$ decreases, larger scales undergo suppression (Eq.~\ref{eq:WDMfsscale1}) and we need correspondingly larger values of $N_{\textrm{ion}}$ which are given by $N_{\rm ion} = \{57,\ 225,\ 861\}$ for decreasing values of the benchmark $m_{\textrm{WDM}}$ values.

\begin{figure}
\centering
\includegraphics[width=1.05\textwidth, angle=0]{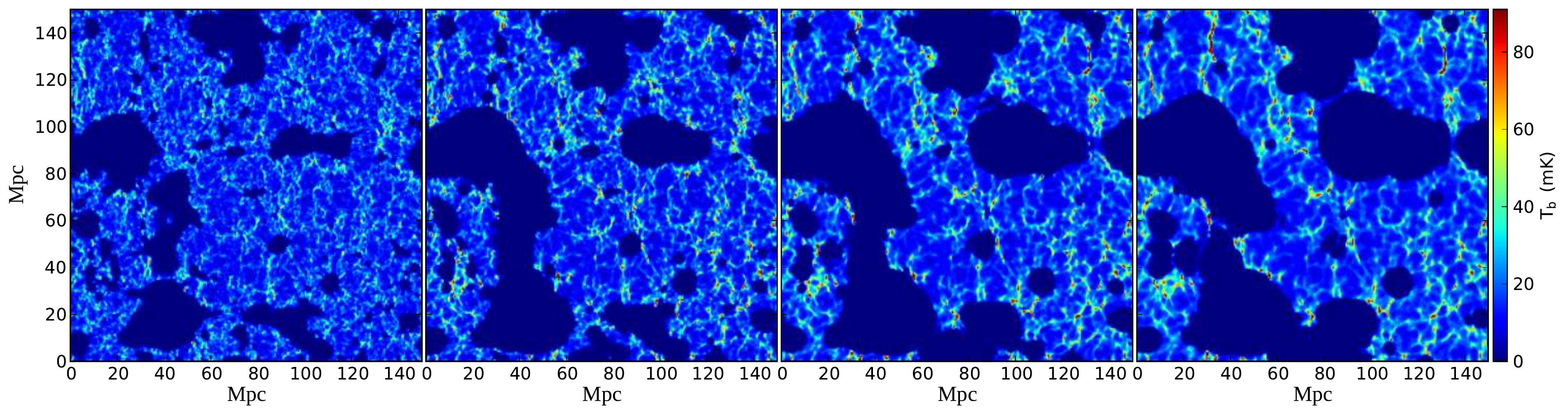}
\put(-409, 105){\textcolor{white}{\bf{{ $\pmb{\Lambda}$CDM}}}}
\put(-338, 105){\textcolor{white}{\bf{{ $\pmb{a_4=0.6\times 10^5 \textrm{ Mpc}^{-1}}$}}}}
\put(-233, 105){\textcolor{white}{\bf{{ $\pmb{a_4=4.2\times 10^5 \textrm{ Mpc}^{-1}}$}}}}
\put(-128, 105){\textcolor{white}{\bf{{ $\pmb{a_4=1.2\times 10^6 \textrm{ Mpc}^{-1}}$}}}}
\caption{Two dimensional sections through the simulated brightness temperature maps of the $\Lambda$CDM model and the three ETHOS benchmark models at $z=8$. From left to right the values of the parameter $N_{\textrm{ion}}$ have been chosen to be $\{23, 100, 301, 721\}$, respectively to ensure $\xb(z=8)=0.5$. As we increase the value of $a_4$ in the ETHOS models, the suppression of low mass halos increases and we need to increase $N_{\textrm{ion}}$ in order to achieve the same level of ionization at $z=8$.}
\label{fig:ETHOS_map}
\end{figure}

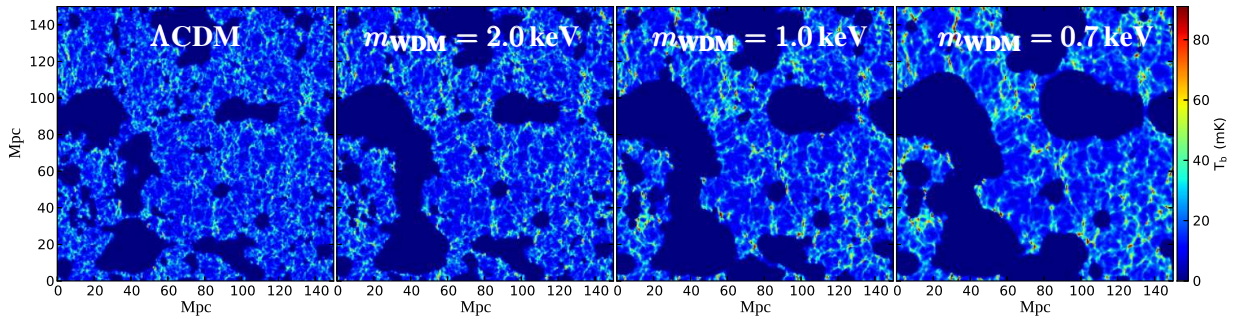
\begin{figure}
\centering
\includegraphics[width=1.05\textwidth, angle=0]{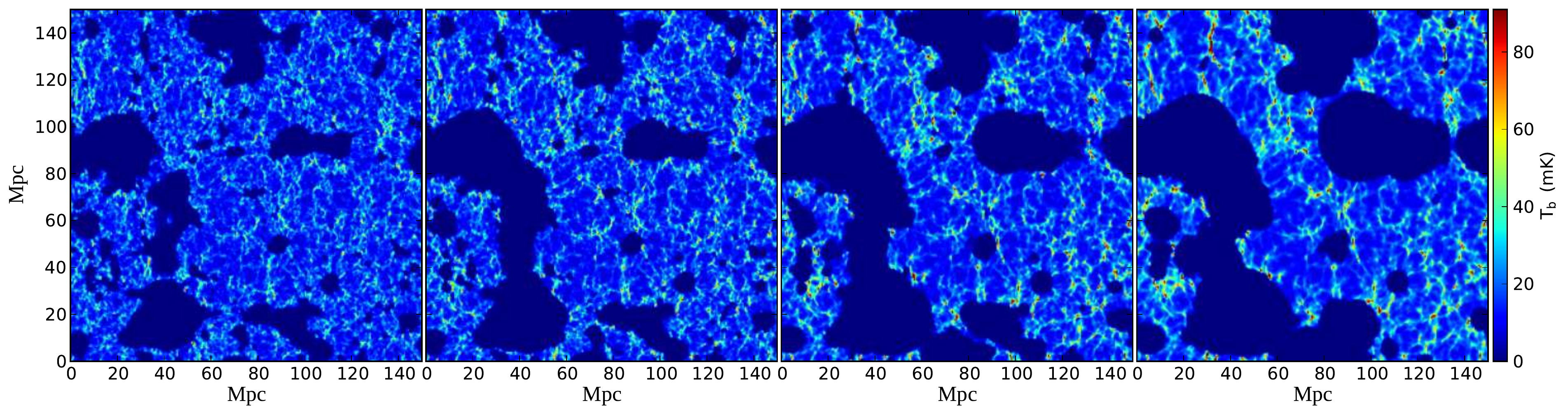}
\put(-409, 105){\textcolor{white}{\bf{{ $\pmb{\Lambda}$CDM}}}}
\put(-329, 105){\textcolor{white}{\bf{{ $\pmb{m_{\textrm{WDM}}=2.0}\,$keV}}}}
\put(-221, 105){\textcolor{white}{\bf{{ $\pmb{m_{\textrm{WDM}}=1.0}\,$keV}}}}
\put(-114, 105){\textcolor{white}{\bf{{ $\pmb{m_{\textrm{WDM}}=0.7}\,$keV}}}}
\caption{Two dimensional sections through the simulated brightness temperature maps of the $\Lambda$CDM model and the three WDM benchmark models at $z=8$. From left to right the values of the parameter $N_{\textrm{ion}}$ have been chosen to be $\{23, 57, 225, 861\}$, respectively to ensure $\xb(z=8)=0.5$. As we decrease the value of $m_\textrm{WDM}$ in the WDM models, the suppression of low mass halos increases and we need to increase $N_{\textrm{ion}}$ in order to achieve the same level of ionization at $z=8$.}
\label{fig:WDM_map}
\end{figure}

We can translate the ionization maps into \HI brightness temperature maps using Eq. 4 of Ref.~\cite{Bharadwaj:2004it}. Fig.~\ref{fig:ETHOS_map} and Fig.~\ref{fig:WDM_map} show two dimensional sections through the simulated \HI brightness temperature \cite{Bharadwaj:2004it} maps of the three ETHOS parameter points and three WDM parameter points along with the $\Lambda$CDM model at $z=8$.

By visual inspection of Fig.~\ref{fig:ETHOS_map} and Fig.~\ref{fig:WDM_map} we can clearly see the main difference between $\Lambda$CDM and ETHOS/WDM models is that the size of the ionized regions is larger in the ETHOS/WDM models. This is a consequence of the fact that the sources require higher values of $N_{\textrm{ion}}$ to achieve the desired level of ionization. This increased value could be assumed to originate form a higher star formation efficiency than the nominal value of $10\%$ that we have assumed. Our method predicts an `inside-out' ionization, where the high density regions are ionized first and the low density region later. As the value of $a_4$ is increased in the ETHOS models or $m_{\textrm{WDM}}$ is decreased in WDM models, the number of low mass halos decreases. This means that the number of ionizing sources in the ETHOS and WDM models is smaller than in the $\Lambda$CDM model.

\subsubsection{Setting a constraint on ETHOS and WDM model parameter space}
To recap our procedure we have used three main steps: (i) we have generated the dark matter distribution at a desired redshift, (ii) we then identified the location and mass of collapsed halos, (iii) we then generated the neutral hydrogen map using an excursion set formalism \cite{Furlanetto:2004nh}. These techniques have been well established in previous work studying the physics of reionization (see for e.g.~\cite{Mondal:2014xma,Mondal:2015oga,Mondal:2016,Mondal:2017}).

We have seen (as shown in Fig.~\ref{fig:ETHOS_map} and Fig.~\ref{fig:WDM_map}) that in order to achieve the same 50\% ionization level at redshift $z=8$, $N_{\rm ion}$ is higher in the ETHOS and WDM models as compared to the $\Lambda$CDM model. If we impose the requirement that values of $N_{\rm ion} \gtrsim 500$ are unphysical, then we can conclude that ETHOS models with $a_4 \gtrsim 1.2\times 10^6 \textrm{ Mpc}^{-1}$ and WDM models with $m_{\textrm{WDM}} \lesssim 0.7\,{\rm keV}$ are excluded. This result is the main contribution of our work\footnote{We note that our WDM constraint is consistent with the result of Ref.~\cite{barkana01} which also tried to examine constraints on WDM scenarios from observables stemming from the epoch of reionization.}.

\subsubsection{Consistency checks of our procedure and subtle issues}
One may ask whether is it justified to neglect the halos of mass $10^8 < M_{\rm halo} < 10^9\, {\rm M_{\odot}}$ in our $N$-body simulation. In a previous work involving two of the current authors~\cite{sarkar16}, higher resolution simulations with the smallest halo masses $\sim 10^8\, {\rm  M_{\odot}}$ were performed in the context of ultra-light axion dark matter models which also resulted in a suppression of small scale structure. The results were found not to change qualitatively upon neglecting halos with masses between $10^8 < M_{\rm halo} < 10^9\, {\rm M_{\odot}}$, since the abundance of smaller halos is suppressed as compared to the $\Lambda$CDM model and the luminosity of ionizing photons is proportional to the mass of the halo. A similar effect can be expected in the ETHOS/WDM models considered in this work. Note that we do not consider the possible impact on reionization of H$_2$-cooled mini-halos.

Another well known issue with $N$-body simulations in non-$\Lambda$CDM models is the presence of spurious halos~\cite{Wang:2007he,Lovell:2013ola,Agarwal:2015iva}, which are artifacts of the naive halo-finding procedure that we have employed. This leads to an over prediction of the number of low mass halos which is visible as an upturn in the halo mass function at low masses~\cite{Schneider:2013ria}. We can see such a feature in our simulations for masses $\lesssim 10^{10}$~${\rm M_{\odot}}$ in Fig.~\ref{fig:mass_func}. In previous studies~\cite{Agarwal:2015iva}, it has been found that these halos have systematically larger spin values, high ellipticity and prolateness, and also display significant deviations from virial equilibrium. These features can be used to identify and remove these halos and obtain the correct halo mass function. This procedure is quite challenging and beyond the scope of this paper and we therefore reserve this for future work. However, we will comment on the expected effect of the removal of the spurious halos on our constraint on $a_4$ and $m_\textrm{WDM}$. Using the halo mass function in Fig.~\ref{fig:mass_func} and the assumed linear relationship between halo mass and the number of ionizing photons per halo (Eq.~\ref{eq:Nion1}),  we can estimate the fraction of ionizing photons that are produced in our simulations from halos of mass $\lesssim 10^{10}$~${\rm M_{\odot}}$ compared to the total number of photons as $\sim$~$40\%$. Thus, even if the true mass function after removal of the spurious halos displays a sharp cut-off below $ 10^{10}$~${\rm M_{\odot}}$, we would expect an approximate increase in the corresponding value of $N_\textrm{ion}$ by $60\%$ to maintain $50\%$ ionization at $z=8$. For the benchmark points that we have considered, this would not change our conclusions about which parameter points are ruled out. If we had scanned over more intermediate parameter points, we would expect that the removal of the spurious halos would have strengthened our constraint slightly.

We have ignored dark matter self-interactions in the ETHOS models when performing our $N$-body simulations. While the DM-DR scattering rate is low enough that it can be neglected, we note that the existence of a light vector mediator ($\phi_\mu$) in ETHOS model 1 could lead to a Sommerfeld enhancement~\cite{Tulin:2012wi} of the DM-DM elastic self-scattering cross-section. In order to have an appreciable effect on virialized halos, it has been estimated that the ratio of energy transfer crosssection to dark matter mass would have to be on the order of $\sim 0.01-1$~cm$^2$/g~\cite{Feng:2009mn,Agrawal:2016quu}. While this could happen for certain choices of the masses and couplings in the ETHOS model, it is typically only important when the mass of the mediator is extremely light as compared to the dark matter mass~\cite{Ibe:2009mk}\footnote{Ref.~\cite{Binder:2017lkj} has considered the effect of Sommerfeld enhancement leading to an epoch of late time reannihilation of the dark matter into dark radiation in the ETHOS 1 model and explored some interesting cosmological signatures.}.

We have also generated ionization maps at different redshifts and after integrating over redshift, we found $\tau_{\textrm{reio}}$ values that lie between $0.046-0.061$ from our simulations for all the benchmark ETHOS and WDM models that we have studied and for the $\Lambda$CDM model point. This serves as a consistency check that our naive criteria of requiring $\xb(z=8)=0.5$ automatically satisfies the optical depth constraint.

Our choice of $50\%$ ionization at $z=8$ was a simplified criteria to satisfy the optical depth constraint from Planck as well as the Gunn-Peterson bound. We present in Table~\ref{tab:xHI} the values of $N_{\textrm{ion}}$ that would be needed if we changed the criteria to $40\%$ or $60\%$ ionization at $z=8$ for the standard $\Lambda$CDM model, as well as for the ETHOS and WDM benchmark model points. Our choice of requiring $N_{\textrm{ion}} \lesssim 500$ as a realistic parameter quantifying the number of ionizing photons per halo would rule out $a_4$ = $1.2 \times 10^6$ Mpc$^{-1}$ and $m_{\textrm{WDM}}$ = 0.7~keV for any of these assumed values of the ionization fraction at $z=8$, while allowing the other benchmark model points. Thus the exclusion limit on the ETHOS and WDM parameters that we have set is robust to moderate changes in the assumed ionization fraction at $z=8$ within the crude resolution of our benchmark parameters.

\begin{table}
\begin{centering}
\begin{tabular}{|c|c|c|c|}
%\backslashbox{Model}{$\xb(z=8)$}
%&\makebox[3em]{ 40\%}&\makebox[3em]{50\%}&\makebox[3em]{60\%}
% \\ \hline\hline

  \hline
  Model/$\xb(z=8)$ & 40\% & 50\% & 60\% \\ \hline \hline
  $\Lambda$CDM &	28	& 24 &	19  \\ \hline
  ETHOS $a_4$ = $0.6 \times 10^5$ Mpc$^{-1}$	 & 121  & 100 & 80\\
  ETHOS $a_4$ = $4.2 \times 10^5$ Mpc$^{-1}$	& 380&	300&	234 \\
  ETHOS $a_4$ = $1.2 \times 10^6$ Mpc$^{-1}$	&955&	721	& 541 \\ \hline
  WDM $m_{\textrm{WDM}}$ = 2.0~keV &	69 &	58 &	47 \\
  WDM $m_{\textrm{WDM}}$= 1.0~keV  &	282	& 226 &	178 \\
  WDM $m_{\textrm{WDM}}$ = 0.7~keV &	1155 &	861 &	645\\ \hline \hline
\end{tabular}
\caption{Table of $N_{\textrm{ion}}$ values for $\Lambda$CDM, ETHOS and WDM benchmark models for different assumptions of the value of the neutral fraction at redshift $z=8$, $\xb(z=8)$. Imposing that values of $N_{\textrm{ion}}>500$ are unphysical would lead us to the same conclusions for the exclusions on the benchmark ETHOS and WDM models for any of the assumed values of $\xb(z=8)$.}
\label{tab:xHI}
\end{centering}
\end{table}

We note that the ability to set a strong bound on ETHOS/WDM models is highly dependent on the assumption that $N_{\textrm{ion}}$ is independent of the halo mass. Cosmological hydrodynamic simulations have indicated that $N_{\textrm{ion}}$ should increase with halo mass, although the exact dependence is still uncertain~\cite{2011MNRAS.410.1703F}. Although the assumption of constant $N_{\textrm{ion}}$ is standard in the literature, it is possible that higher mass halos could be responsible for most of the reionization (for example, if $N_{\textrm{ion}}$ is a rapidly increasing function of halo mass~\cite{doi:10.1111/j.1365-2966.2007.12279.x, 2016MNRAS.456.2080M}). In such a case, the suppression of low mass halos will not have a noticeable impact on the reionization history and we will not be able to set a strong constraint on the ETHOS and WDM models.

Another subtle but related issue is that the dependence of $N_{\textrm{ion}}$ on halo mass could also be related to the level of suppression of small scale structure in ETHOS/WDM models~\cite{Bose:2016irl,Bose:2016hlz,Lovell:2017eec}. It was noticed in hydrodynamic simulations that halos with masses below the suppression scale tend to have more ionizing UV photons than similar mass halos in the standard $\Lambda$CDM cosmology at $z>7$. The increased number of UV photons in the ETHOS/WDM models has been interpreted as being caused due to the increased formation of starburst galaxies created in the merger of heavy galaxies in these cosmologies. This is in contrast to $\Lambda$CDM simulations, where it was observed that the formation of starburst galaxies is suppressed by supernova feedback from low mass galaxies, which drives the star forming gas out of merging galaxies. The expected increase in UV photons from intermediate mass halos (just below the cut-off) in ETHOS and WDM models was found to partially compensate for the loss of ionizing photons due to the suppression of the number of smaller galaxies. Ref.~\cite{Lovell:2017eec} found only a $10\%$ decrease in optical depth in their benchmark ETHOS model due to the overall decrease in the number of ionizing photons.

\section{Future outlook of reionization constraints from 21-cm measurements}
\label{sec:21cm}

Current and future observations of the 21-cm line of neutral Hydrogen could provide us with a much more precise probe of the epoch of reionization. Observations of the 21-cm line could probe the \HI power spectrum at different redshifts and in addition they could also be used to measure the redshift dependence of $\xb(z)$ (or the so called global history of reionization). There is a considerable effort already underway to detect the epoch-of-reionization 21-cm brightness temperature fluctuation power spectra through radio interferometry observatories such as GMRT, LOFAR, MWA and PAPER~\cite{2013MNRAS.433..639P,2013A&A...550A.136Y,2014PhRvD..89b3002D,2015ApJ...809...61A}. Apart from these first-generation radio interferometers, the detection of this signal is also one of the key science goals of future radio telescopes such as SKA and HERA~\cite{2013ExA....36..235M,2009astro2010S..82F}. To complement this effort, there are also experiments seeking to measure the global 21-cm signal, such as EDGES, SARAS, and DARE~\cite{2010Natur.468..796B,2017ApJ...845L..12S,2012AdSpR..49..433B}. These measurements could be used to set tighter constraints on the ETHOS and WDM model parameters and self-interacting dark matter scenarios in general.

\begin{figure}
\psfrag{k3 P(k)/2 pi2}[c][c][1][0]{$\Delta_{\rm b}^2\, (k)$ ($\rm mK)^2$}
\psfrag{k (Mpc-1)}[c][c][1][0]{$k~({\rm Mpc}^{-1})$}
\psfrag{WDM m=2.4keV}[c][c][1][0]{\footnotesize{WDM $m_{\textrm{WDM}}=2.4$keV~~~~}}
\psfrag{CDM}[c][c][1][0]{{$\Lambda$CDM}\, \,}
\centering
\includegraphics[width=1.05\textwidth]{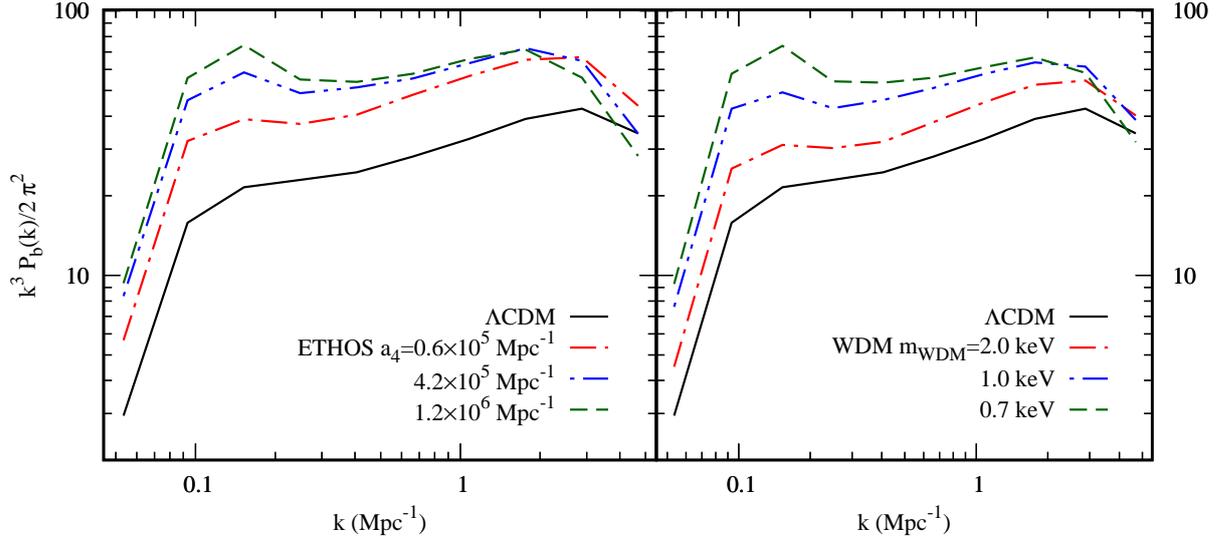}
\caption{The figures above show the brightness temperature power spectrum $\Delta_{\rm b}^2\, (k)$ of the \HI field as a function of $k$ at $z=8$ for our three different benchmark ETHOS and WDM models. The black solid curves shows the $\Lambda$CDM \HI brightness power spectrum. Models with fewer low mass halos (larger $a_4$ or smaller $m_\textrm{WDM}$) have fewer, but more intense ionizing sources. The localization of ionized bubbles leads to larger \HI regions which is reflected in the increased brightness power spectra on large scales ($\sim$~$1-10$~Mpc) in the figure above.}
\label{fig:pk_ionz}
\end{figure}

In Fig.~\ref{fig:pk_ionz} we show our prediction for the brightness temperature fluctuation power spectra $\Delta_{\rm b}^2\, (k)=k^3 P_b\,(k)/2 {\rm \pi}^2$ of the \HI field at $z=8$ for the ETHOS and WDM benchmark models that we have considered here along with the $\Lambda$CDM  model, for a fixed ionization fraction $\xb=0.5$. We find that the power for ETHOS and WDM models is greater than the $\Lambda$CDM model over a large range of scales $0.05 \, {\rm Mpc^{-1}} < k < 4 \, {\rm Mpc^{-1}}$. This feature is to be expected from our discussion in the last section and by reexamining Fig.~\ref{fig:ETHOS_map} and Fig.~\ref{fig:WDM_map}. The reason is as follows: With increasing suppression of smaller halos in the ETHOS and WDM models, large halos must contribute more to the ionization of the universe which is assumed to arise from a larger $N_{\textrm{ion}}$ parameter (more ionizing photons per unit halo mass). As reionization proceeds, larger regions of ionizing bubbles form around halos as compared to the standard $\Lambda$CDM model, where $N_{\textrm{ion}}$ is lower. Since the only constraint we are imposing is that the ionization fraction at $z=8$ is $50\%$, the \emph{neutral} \HI regions are also contiguously larger and relatively undisturbed since this condition can be satisfied with the larger ionized bubbles that sit outside the \HI regions\footnote{We have also checked that the 21-cm brightness spectrum that we have obtained is not altered significantly by changing our criteria on the neutral \HI fraction $\xb=0.5$ to $\xb=0.4$ or $\xb=0.6$ at $z=8$.}. Thus, we expect an enhanced 21-cm brightness temperature on large scales in the WDM and ETHOS scenarios. The relationship between larger ionized bubbles and the increased 21-cm brightness power spectrum on large scales has been shown analytically \cite{Furlanetto:2004nh} and is consistent with the results of numerical simulations \cite{Lidz:2007az}.

If deviations from the standard $\Lambda$CDM scenarios are observed either in the global history of reionization or in the 21-cm brightness power spectrum, confirming them as signs of non-standard dark sector interactions is trickier due to the poorly known astrophysics of reionization. In order to make progress, it will be important to be able to pin down the astrophysical uncertainties on the parameters that determine $N_{\textrm{ion}}$. One way this could be achieved, is through direct observations and subsequent studies of early galaxies responsible for reionizing the universe~\cite{2017NatAs...1E..91H, 2006MNRAS.367L..11S, 2013MNRAS.431..383Y} or observation of Pop III stars \cite{2015MNRAS.450.2506V, 2016ApJ...820...59K}.

%%%%%%%%%%%%%%%%%%%%%%%%%%%%%%%%%%%%%%%%%%%%%%%%%%%%%%%%%%%%%%%%%%%%%%%%%%%%%%%%%%%%%%%%%%%%
\section{Summary and conclusions}
\label{sec:conclusions}
In this work, we have examined the impact of dark matter-dark radiation interactions on the epoch of reionization. We worked with the generalized ETHOS framework which allowed us to parameterize DM-DR interactions in terms of a few coefficients of a redshift series expansion of the relevant collision terms in the Boltzmann equations. We chose a particular model (ETHOS 1) as a case study which involved a sterile neutrino dark radiation component interacting with DM through a massive vector mediator. We gave qualitative arguments to show that DM-DR interactions lead to a suppression of small scale structure (relative to the standard non-interacting DM scenario) and we presented analytical formulae that can be generalized to other ETHOS models for the suppression scale in the linear matter power spectrum. We also drew comparisons with warm dark matter scenarios where similar suppression of structure on small scales is expected. We then computed the numerical linear power spectra in both ETHOS and WDM models and showed that they exhibit the expected suppression on small scales, but DM-DR interactions also led to prominent acoustic oscillation features in the matter power spectrum of ETHOS models.

We then showed that non-linearities smooth out the acoustic oscillation features and therefore the late time power spectra of ETHOS and WDM models (for adjusted parameter choices) can look similar. In particular we used $N$-body simulations to find the power spectrum at redshift $z =8$, and we showed that the suppression of the power spectra relative to the standard $\Lambda$CDM picture is a feature that persists even when non-linearities are taken into account. Consequently, we showed that this led to a suppression of low mass halos.

We then discussed the impact on reionization due to the suppression of low mass halos. This connection was established by defining a parameter $N_\textrm{ion}$ which can be thought of as the number of ionizing photons escaping the halo per unit mass, where we used hydrogen mass as the reference unit mass. Despite the large astrophysical uncertainties on the value of $N_\textrm{ion}$, we concluded that one can safely claim that $N_\textrm{ion} \lesssim 500$.

In order to satisfy the optical depth observations from Planck and the Gunn-Peterson bound, we demanded that a successful reionization history in any DM scenario must achieve 50\% reionization by redshift $z=8$. In the ETHOS and WDM models, as the small halos get more and more suppressed, one needs to increase the value of $N_\textrm{ion}$ in order to achieve the same level of reionization. However, if the suppression of small scale structure is too much, this would require unphysical values of $N_\textrm{ion} > 500$. We therefore used this as a constraint to set bounds on the ETHOS and WDM model parameters.

We found that for ETHOS model 1 the parameter range $a_4 \gtrsim 1.2\times 10^6 \textrm{ Mpc}^{-1}$ (defined in Sec.~\ref{sec:ethosmodels}) and for WDM models the parameter range $m_{\textrm{WDM}} \lesssim 0.7\,{\rm keV}$ are excluded.

It is interesting to note that the bounds on the WDM that we have obtained are comparable to (although weaker than) bounds set by Lyman-$\alpha$ measurements around $z \sim 3$ which give $m_\textrm{WDM} \gtrsim 5$~KeV~\cite{Murgia:2017lwo}. We have also argued how future 21-cm measurements of the global history of reionization and the \HI power spectrum could help set improved bounds on DM-DR interactions. Confirming a deviation in any epoch of reionization observable as a sign of secret dark sector interactions is much more difficult. This would require a better understanding of the astrophysical processes that drive reionization. This could happen through direct observations of early galaxies or Pop III stars.

%However, our bound is complementary to the Lyman-$\alpha$ bounds, since we are setting the constraint at higher redshifts.

In conclusion, our work has demonstrated that the epoch of reionization can provide an interesting probe of DM-DR interactions. By working in the generalized ETHOS framework, we have created a path from converting the particle physics parameters to a suppression scale in the linear matter power spectra, and then set a bound on the amount of suppression in the matter power spectrum from demanding a successful reionization history.

While the ETHOS constraint we have derived is specific to ETHOS model 1, our general arguments and use of the ETHOS framework allow us to potentially generalize the reionization constraints to \emph{any} kinds of dark sector with DM-DR interactions. While the connection between the particle physics parameters and the suppression scale of the linear matter power spectrum have a simple analytical relationship, the reionization constraints require numerical analysis. We plan to develop a simplified version of this constraint (based on the a scan over several suppressed linear matter power spectra) in a future work, so that the EoR bounds can be used by model builders to quickly assess the EoR constraints on self-interacting dark sector models.

\section{Acknowledgments}
We would like to thank the anonymous referee for some valuable comments and suggestions to improve the manuscript. We would like to thank Tirthankar Roy Choudhury, Francis-Yan Cyr-Racine, Takeshi Kobayashi, Aseem Paranjape and Matteo Viel for helpful conversations and correspondence. S.~D. thanks Prof. Dominik Schwarz at Bielefeld Physics Dept for hosting him in summer 2017 under a DFG initiation grant, where considerable work was done. V.~R. would like to thank ICTP, Trieste where a part of this work was completed. V.~R. is supported by a DST-SERB Early Career Research Award (ECR/2017/000040) and an IITB-IRCC seed grant.

\appendix

\section{The Jeans scale in ETHOS and WDM models}
\label{sec:appendix}

\subsection{Qualitative understanding of evolution of linear density perturbations in ETHOS models}
We will first attempt to understand the impact of ETHOS parameters on the linear matter power spectrum. The key criteria is to determine the evolution of the Jeans scale in different cosmological epochs~\cite{2010gfe..book.....M}. Density perturbations on scales smaller than the Jeans scale are prevented from collapsing and thus the linear matter power spectrum on such small scales would be suppressed. We will attempt to map the ETHOS parameters (and hence the particle physics parameters) to the evolving Jeans scale. Since damping of the power spectrum on very large scales would be in conflict with observations, we will focus on a range of ETHOS parameters that lead to a maximum comoving Jeans scale of at most $\lambda_J \sim 1$~Mpc.

There are several interesting features that arise in the matter power spectrum due to the presence of DM-DR scattering.
\begin{itemize}
\item In the early universe when DM and DR are tightly coupled, the collapse of DM density perturbations is prevented by the pressure created by the DR fluid. Dark acoustic oscillations (DAO), similar to baryon acoustic oscillations are set up due to the interplay between gravitational attraction and pressure generated by DR. The Jeans scale in this epoch is given by the sound horizon.
\item While DM and DR are coupled, collisions will lead to a diffusion (Silk) damping of the oscillation power spectrum.
\item At later times after DM and DR decouple from each other, free-streaming of the DM prevents collapse on small scales. The Jeans scale in this era is just given by the free-streaming length of dark matter. We thus expect an overall attenuation of the power spectrum on scales smaller than the Jeans scale due to free-streaming of dark matter in this epoch.
\end{itemize}

The oscillatory and damping features are clearly visible when one numerically solves the full set of coupled DM-DR-metric perturbation evolution equations. However, we would like to qualitatively analyze how the scales of DAO, damping and free-streaming depend on the DM-DR interaction strengths. To study these effects on the DM power spectrum, we need to identify the key transitions between the epochs of interest.

\begin{figure}
 \includegraphics[width=\linewidth]{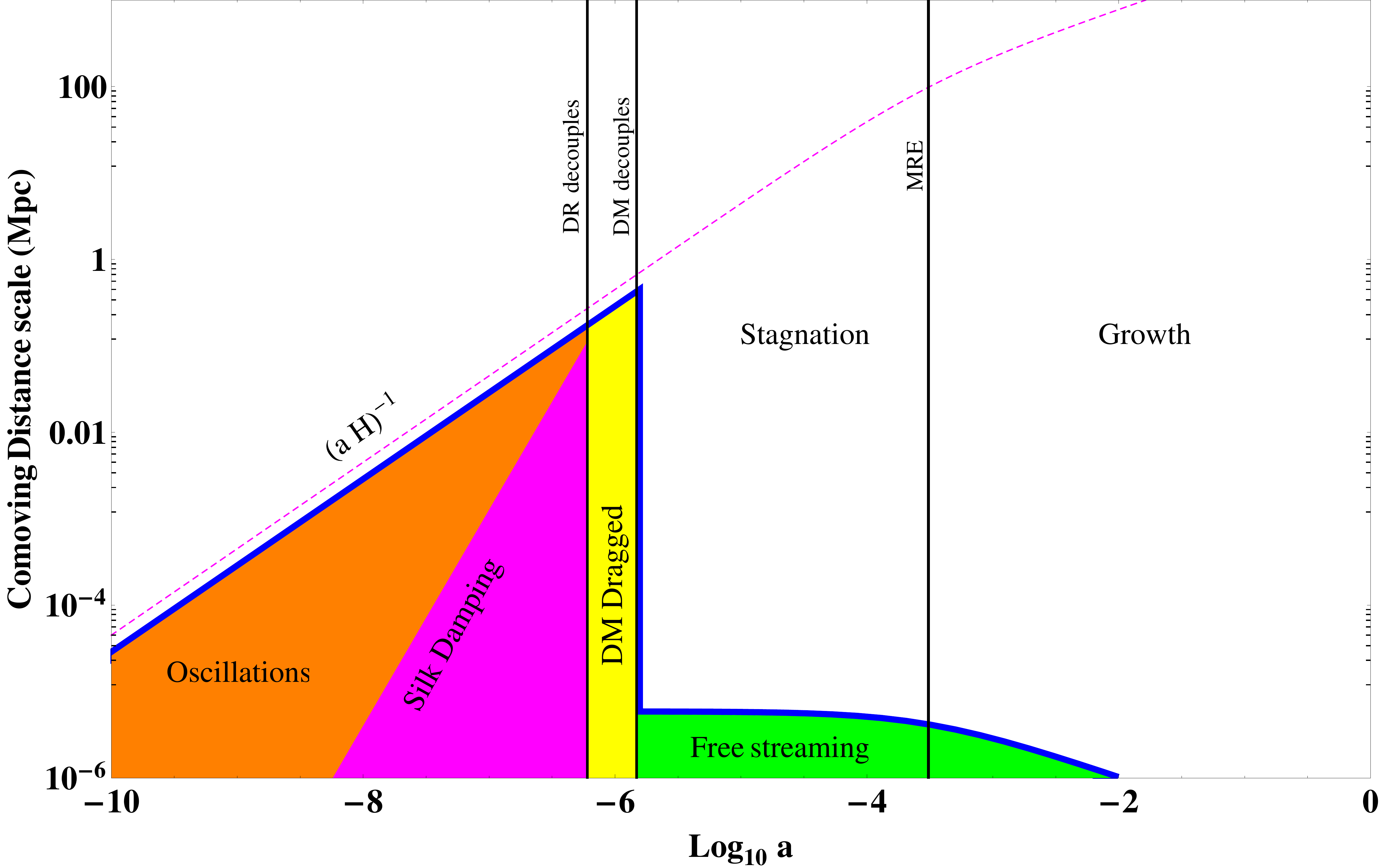}
\caption{The evolution of the Jeans length scale (blue solid line) in the early universe in ETHOS model 1 with $a_4=0.6\times10^5 \textrm{ Mpc}^{-1}$ as a function of the FRW scale factor $a$. The epochs of DR decoupling, DM decoupling and matter-radiation equality are clearly marked. The behavior of the modes smaller than the Jeans scale is initially suppressed due to acoustic oscillations and later due to non-relativistic free streaming. The maximum damping scale is set by the Jeans length at DM decoupling and modes below this scale will be exponentially damped. Also shown for reference is the comoving Hubble length (dashed magenta line).}
\label{fig:jeanslength}
\end{figure}

Fig.~\ref{fig:jeanslength} shows the evolution of the Jeans scale as a function of the FRW scale factor ($a$) for ETHOS model 1 using a benchmark parameter $a_4 = 0.6 \times 10^5 \textrm{ Mpc}^{-1}$. Also shown is the comoving Hubble scale $(a H)^{-1}$ as a function of the scale factor. The key transitions in the behavior of the evolution of the DM density perturbations occur at four distinct times (three of which are shown by the vertical lines in Fig.~\ref{fig:jeanslength}) which we discuss below.

\begin{itemize}
\item The first critical transition in behavior occurs when DM becomes non-relativistic. We assume this happens very early on while DM is still coupled to the DR fluid. This transition is assumed to happen very early and is \emph{not} shown in the figure.
\item Eventually DM and DR interactions fall out of equilibrium. There are actually two decoupling times of interest, the first is when DR decouples and begins to free stream. This occurs when $-\dot{\kappa}_{\textrm{DR}-\textrm{DM}} \simeq a H$. The second is when DM stops scattering off of the free-streaming DR fluid, when $-\dot{\kappa}_{\chi} \simeq a H$. We denote the scale factor at these time as $a^{\textrm{DR}}_\textrm{dec}$ and $a^{\textrm{DM}}_\textrm{dec}$ respectively. Note that we are making a simplifying assumption of \emph{instantaneous} decoupling. Both these decoupling transition times are shown in Fig.~\ref{fig:jeanslength}. We will assume that DM decouples\footnote{Henceforth, whenever we use the term ``decoupling'', we use it to mean kinetic decoupling.} deep in the radiation epoch. Since we have already assumed it is non-relativistic, it thus survives as a cold relic.
\item The epoch of matter radiation equality (MRE) for the universe is also shown in this figure. We denote the scale factor at this time as $a_\textrm{MRE}$. The exact value of $a_\textrm{MRE}$ has a slight dependence on the value of $\Omega_\textrm{DR}$ but is mostly unchanged from the standard $\Lambda$CDM picture, $a_\textrm{MRE} = \frac{ \Omega_\textrm{r}h^2}{\Omega_\textrm{m}h^2} \approx 3 \times 10^{-4}$. Here, as usual, $\Omega_\textrm{r}$ and $\Omega_\textrm{m}$ denote the total radiation and matter density fractions today.
\end{itemize}

We can estimate the decoupling scale factors at the two decoupling transitions in any given ETHOS model as,
\begin{align}
  a^\textrm{DR}_\textrm{dec} &=  \left ( \frac{\Omega_\textrm{\tiny DM} h^2}{\Omega_r^{1/2}}  \frac{a_n}{H_0 (1+z_d)^n} \right )^{1/(n-1)}, \nonumber \\
  a^\textrm{DM}_\textrm{dec} &=  \left ( \frac{\Omega_\textrm{\tiny DR} h^2}{\Omega_r^{1/2} } \frac{4}{3} \frac{a_n}{H_0 (1+z_d)^n} \right )^{1/n}.
\end{align}
Here, we have assumed that $\dot{\kappa}_\chi \propto a_n (1+z)^{n+1}$ is a model dependent redshift scaling of the DM mean free path. For the specific ETHOS model considered in Sec.~\ref{sec:ethosmodels} (where $n=4$), we find the following dependence of the scale factor at decoupling on the ETHOS parameters,
\begin{align}
  a^\textrm{DR, mod1}_\textrm{dec} &=  6.91 \times 10^{-7}   \left(  \frac{a_4}{ 0.6 \times 10^5 \textrm{ Mpc}^{-1}} \right )^{1/3}, \nonumber \\
  a^\textrm{DM, mod1}_\textrm{dec} &=  1.48 \times 10^{-6}   \left(  \frac{a_4}{ 0.6 \times 10^5 \textrm{ Mpc}^{-1}} \right )^{1/4}.
\end{align}

As noted above, we have assumed instantaneous decoupling at a given redshift. However, this assumption depends on the width of the redshift dependence of the \textit{visibility function} $-\dot{\kappa}_{\chi} e^{- \kappa_{\chi}}$. The visibility function can be roughly thought of as a probability distribution function for the redshift of last scattering between DM and DR. Ref.~\cite{Cyr-Racine:2015ihg} showed that as $n$ increases, the visibility function becomes narrower and the instantaneous decoupling approximation is justified. A second assumption that we make is that both these decouplings take place well before MRE in order to avoid major changes to the matter power spectrum on large scales on the order of the comoving Hubble horizon size at MRE ($\sim$ 100 Mpc)\footnote{Decoupling post MRE would lead to oscillations of the power spectrum on very large scales, in a scenario akin to a baryon dominated universe.}.

We are now ready to understand the qualitative features that are expected to show up in the matter power spectrum and the scales at which they would operate. We will discuss the Jeans scale in each of four distinct epochs.

\textbf{Before DR decoupling (tightly coupled epoch):} In the very early universe, before DR decoupling has taken place, the DM-DR fluid oscillates with a sound-speed $c_s$ which is given at conformal time ($\tau$) by,
\begin{equation}
\label{eq:soundspeed}
c^2_s(\tau) = \frac{1}{3} \left ( \frac{1}{1 +   R_d}     \right )  = \frac{1}{3} \left ( 1 +   \left ( \frac{3}{4}  \frac{ \Omega_\textrm{\tiny DM} h^2  }{\Omega_\textrm{\tiny DR} h^2} a \right )    \right )^{-1} .
\end{equation}
In analogy with the coupled baryon-photon fluid, we have defined the parameter $R_d$ which is related to the DM and DR densities as $R_d =   \frac{3}{4} \frac{ \rho_\textrm{\tiny DM}   }{\rho_\textrm{\tiny DR}}  = \frac{3}{4} \frac{ \Omega_\textrm{DM} h^2  }{\Omega_\textrm{DR} h^2} a $.

The tightly-coupled DM-DR fluid oscillates as a single fluid until DR decoupling. The instantaneous sound horizon before DR decoupling as a function of conformal time $(\tau = \int dt/a(t))$ is given by,
\begin{equation}
r_s(\tau) \simeq  \frac{c_s(\tau)}{a H} .
\end{equation}
The sound horizon corresponds to the Jeans scale in this epoch (See Fig.~\ref{fig:jeanslength}).
This gives us a maximum sound horizon scale at DR decoupling of,
\begin{equation}
\label{eq:maxSH}
r^\textrm{max}_s \equiv r_s(\tau^{\textrm{DR}}_\textrm{dec}) \simeq \frac{1}{\sqrt{3}} \frac{a^{\textrm{DR}}_\textrm{dec}}{H_0 \Omega^{1/2}_r} \simeq  0.18  \left ( \frac{a^{\textrm{DR}}_\textrm{dec}} {6.91 \times 10^{-7}} \right ) \textrm{ Mpc} \simeq 0.18  \left ( \frac{a_4}{0.6 \times 10^{5} \textrm{ Mpc}^{-1}} \right )^{1/3} \textrm{ Mpc} ,
\end{equation}
where the last equality holds specifically for ETHOS model 1. Since inflation only excites the cosine modes, this leads to peaks in the linear matter power spectrum at wavenumbers $k_p =  \frac{m \pi}{r^\textrm{max}_s}$, where $m = 1, 3, 5$ etc.

During this epoch, collisional or diffusion damping of modes below the Silk-damping scale $r_{\textrm{SD}}$ will occur. The damping scale is given in the tightly-coupled regime by~\cite{Cyr-Racine:2015ihg},
\begin{equation}
\label{eq:silkdamping}
r_{\textrm{SD}}(\tau) = \pi \left(-1/6 \int_0^\tau \frac{d\tau^\prime}{\dot{\kappa}_{\chi} + \dot{\kappa}_{\textrm{DR-DM}}} \left [ \frac{\dot{\kappa}_{\textrm{DR-DM}} }{\dot{\kappa}_{\chi} + \dot{\kappa}_{\textrm{DR-DM}}} + \frac{ 4 \dot{\kappa}_{\chi} }{5(\alpha_2 \dot{\kappa}_{\textrm{DR-DM}}  )} \right ] \right )^{1/2}.
\end{equation}
Damping will take place on all scales smaller than the Silk damping scale as long as DM and DR are kinetically coupled. The maximum relevant damping scale is thus given by $k_{\textrm{SD}}(\tau^{\textrm{DR}}_\textrm{dec})$, where we evaluate the damping scale at DR decoupling. If we make the reasonable approximation that decoupling occurs when the DR density is still dominant over the DM density, then we can approximate $|\dot{\kappa}_{\textrm{DR-DM}}| \ll |\dot{\kappa}_{\chi}|$ in Eq.~\ref{eq:silkdamping}. This simplifies the expression for the maximum damping scale to,
\begin{equation}
r_{\textrm{SD}}(\tau^{\textrm{DR}}_\textrm{dec}) = \pi \left (   \frac{ (1 + z_D)^n}{6 H_0 \Omega_r^{1/2} (\Omega_\chi h^2)  a_n} \left (  \frac{(a^{\textrm{DR}}_\textrm{dec})^{n+1} }{n+1} \frac{4}{5 \alpha_2}  \right)      \right )^{1/2}.
\end{equation}

Thus for ETHOS model 1 we have,
\begin{align}
\label{eq:SDETHOS1}
r^{\textrm{mod1}}_{\textrm{SD}}(\tau^{\textrm{DR}}_\textrm{dec}) & =   0.13  \left ( \frac{a^\textrm{DR}_\textrm{dec} }{6.91 \times 10^{-7}} \right )^{5/2}  \left (  \frac{0.6\times 10^{5} \textrm{ Mpc}^{-1}   }{ a_4}         \right )^{1/2} \textrm{ Mpc}, \\
&= 0.13   \left (  \frac{ a_4}{0.6\times 10^{5} \textrm{ Mpc}^{-1}   }         \right )^{1/3} \textrm{ Mpc}.
\end{align}

%\blue{Change DR drag to DM drag everywhere}
\textbf{Post DR decoupling but before DM decoupling (DM drag epoch):} Once DR decouples it will free stream, but since the DM continues to be coupled to the dark radiation it will get dragged along with the DR. A full treatment of the Jeans scale is complicated by the ``slip'' between DM and DR. For simplicity, we will assume that DR free streams with an RMS speed $c/\sqrt{3}$ and that the DM gets dragged along without slipping. This assumption will result in us slightly overestimating the relevant Jeans scale which is then given by,
% \blue{This results in an overestimate of the Jeans scale which we can use as an upper bound on the maximum}
\begin{equation}
\lambda_\textrm{fs} (\tau) = \frac{c}{\sqrt{3}}\frac{1}{aH}
\end{equation}
between the epochs of decoupling of DR and DM.

The Jeans scale at the end of this epoch is thus given by,
\begin{equation}
\label{eq:maxjeans}
\lambda_\textrm{fs}  \simeq \frac{1}{\sqrt{3}} \frac{a^{\textrm{\tiny DM}}_\textrm{dec}}{H_0 \Omega^{1/2}_r} \simeq  0.39  \left ( \frac{a^{\textrm{\tiny DM}}_\textrm{dec}} {1.48 \times 10^{-6}} \right ) \textrm{ Mpc} \simeq 0.39 \left ( \frac{a_4}{0.6 \times 10^{5} \textrm{ Mpc}^{-1}} \right )^{1/4} \textrm{ Mpc}.
\end{equation}
Again, the last equality is valid only for ETHOS model 1.

\textbf{Post DM decoupling (DM free streaming epoch):} The next interesting transition is when DM decouples from the free streaming DR. There is a sudden drop in the speed of the DM and it will free stream with a non-relativistic speed given by,
\begin{equation}
c_\chi \simeq \sqrt{\frac{5}{3}\frac{T_\chi}{m_\chi} }.
\end{equation}
Here $T_\chi$ is the effective DM temperature which falls as $\sim 1/a^2$ in this epoch. Density perturbations on scales larger than the free streaming scale will not begin to grow till MRE, rather they will stagnate in this epoch (Meszaros effect~\cite{1974A&A....37..225M}). However, scales smaller than the free streaming scale will be strongly damped. We can find the free streaming length in this epoch as,
\begin{equation}
\label{eq:fs2}
\lambda_\textrm{fs} (\tau) = \frac{c_\chi}{a H} \simeq \frac{v^{\textrm{\tiny DM}}_{\textrm{dec}}}{H_0 \Omega_r^{1/2}}.
\end{equation}

Here, $v^{\textrm{\tiny DM}}_{\textrm{dec}}$ is the velocity of the DM component at the time of decoupling from dark radiation. The last equality is true only deep in the radiation dominated era, and leads to a prediction of a constant Jeans scale in this epoch.
This free streaming scale is highly suppressed compared to the Jeans scale prior to DM decoupling.

\textbf{Post MRE:}
Post MRE dark matter continues to free stream non-relativistically. The Jeans scale in this epoch drops as $a^{-1/2}$, due to the slower growth of the Hubble length scale in the matter dominated era. Density fluctuations on scales larger than the Jeans scale can now experience power law growth.

\vspace{2mm}
Thus in summary, two key features remain imprinted on the linear matter power spectrum post MRE. Firstly, the dominant damping is due to the drag experienced by DM while it is still coupled to the free-streaming DR. Scales smaller than the maximum free streaming length (Eq.~\ref{eq:maxjeans}) are damped in the linear matter power spectrum. Secondly, the imprint of oscillations as a remnant of the DM and DR interactions will remain in the linear matter power spectrum, with peaks visible on scales corresponding to $k_p$ (see the discussion below Eq.~\ref{eq:maxSH}).

The suppression of the power spectrum on small scales will lead to a suppression of low mass halos with masses smaller than the Jeans mass ($M_J$) which is given by,
\begin{equation}
M_J = \frac{4\pi}{3} \rho_m R_J^3 \simeq  10^{10} M_\odot \left( \frac{R_J}{0.39 \textrm{ Mpc}} \right )^3 = 10^{10} M_\odot \left ( \frac{a_4}{0.6 \times 10^{5} \textrm{ Mpc}^{-1}} \right )^{3/4}.
\end{equation}
Here, $M_\odot$ denotes solar mass, $R_J$ is the relevant Jeans radius and $\rho_m$ is the total matter density (DM and baryons). In the last equality we have set $R_J$ to the maximum free-streaming length of DM in ETHOS model 1 (Eq.~\ref{eq:maxjeans}).

\vspace{2mm}

We note below a few key differences between DR/DM decoupling in the scenario that we are considering here and conventional baryon-photon decoupling.
\begin{itemize}
\item Baryon photon decoupling occurs post MRE and therefore affects larger scales than those of DR/DM oscillations.
\item Baryons and photons decouple from each other at epochs which are relatively nearby in redshift because of the phase transition at recombination. There is no such phase transition in the simple dark sectors that we are considering and therefore the decoupling of DR and DM are much more widely separated in redshift. %\blue{6 x10^-7} and {1.4 times 10^-6} respectively for our benchmark point
\item Based on the previous point, we thus have an extended phase of evolution of the DM density perturbations between the epochs of decoupling of DR and DM. During this epoch DM is dragged along with the free streaming dark radiation.
\end{itemize}

We note that the case of a dark ``recombination'' where dark atoms are formed and the consequent impact on structure formation was considered in Ref.~\cite{Cyr-Racine:2013fsa}. In such a scenario, behavior which is more similar to the baryon-photon acoustic oscillations is expected in the dark sector.

\subsection{Difference with WDM}
It is natural to compare the ETHOS models that we have considered in this section to warm dark matter (WDM) scenarios, where one expects a damping of the power spectrum due to free streaming of relativistic dark matter. We briefly review the phenomenology of WDM scenarios and contrast them with the ETHOS scenarios discussed previously.

WDM scenarios have a dark sector consisting of a light (typically keV scale) DM particle which is assumed to decouple early from the thermal bath while it is still relativistic, i.e. when the temperature of the dark matter is much greater than its mass $(m_{\textrm{WDM}})$. Here, we assume for simplicity that the WDM is coupled to the standard model plasma (visible photons) which is at a temperature $T$. The generalization to the case where the dark sector has a dark radiation component is a trivial extension of this scenario\footnote{For all effective purposes, if $\xi$ parameterizes the ratio of the dark sector radiation bath to the visible sector temperature, then one can derive identical formulae for the relevant cosmological scales to those in the text with the replacement $m_{\textrm{WDM}} \rightarrow m_{\textrm{WDM}}/\xi$.}.

Post-decoupling, the DM will free stream with an RMS speed $c/\sqrt{3}$, until the effective DM ``temperature'' drops below the DM mass. We denote this temperature as $T_\textrm{NR}$ (we approximate this condition by setting $T_\textrm{NR} = m_\textrm{WDM}/5$). Thus there are two critical transitions of interest that characterize WDM scenarios: 1) the decoupling transition (we assume the scale factor at this epoch is very small and can be set to zero for our analysis) and 2) the onset of non-relativistic behavior (we assume that this happens when the scale factor of the universe is $a_\textrm{NR}$ and it occurs in the radiation dominated era).

\begin{figure}
 \includegraphics[width=\linewidth]{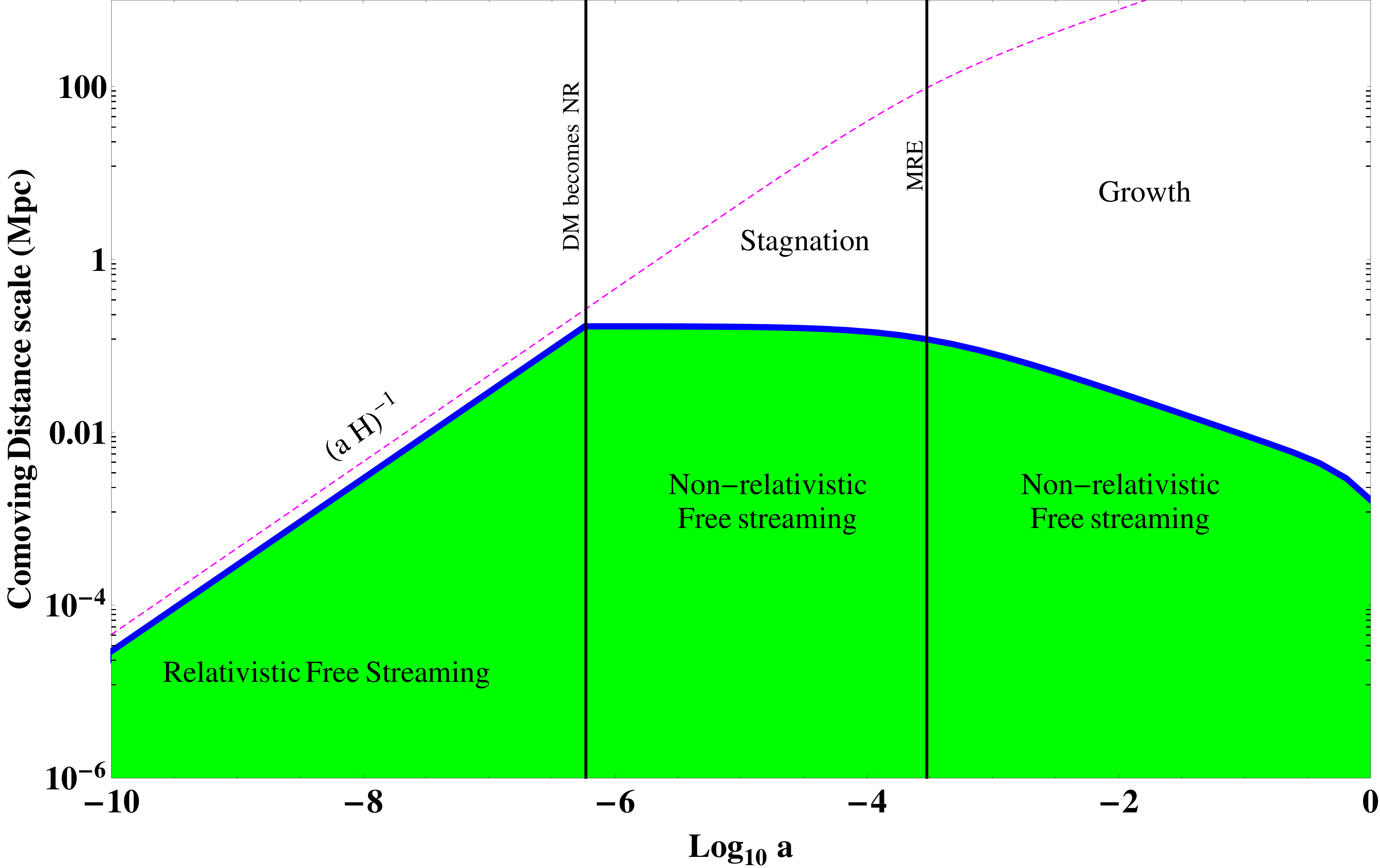}
\caption{The evolution of the Jeans length scale (blue solid line) in the early universe in a WDM model with $m_{\textrm{WDM}}=2 \textrm{ keV}$ as a function of the FRW scale factor $a$. The epochs of DM becoming non-relativistic and matter-radiation equality (MRE) are clearly marked. The growth of the modes smaller than the Jeans scale is initially suppressed due to relativistic and later due to non-relativistic free streaming of dark matter. The maximum damping scale is set by the Jeans length at the onset of non-relativistic free-streaming and modes below this scale will be exponentially damped. Also shown for reference is the comoving Hubble length (dashed magenta line).}
\label{fig:WDMjeanslength}
\end{figure}

Even after DM decouples, it will continue to have an effective thermal temperature that characterizes its phase space distribution. As long as the DM is relativistic, this temperature falls off as $\sim 1/a$. Once the DM becomes non-relativistic, the effective temperature falls off as $\sim 1/a^2$. The scale factor at the onset of the non-relativistic epoch is given by,
\begin{equation}
a_\textrm{NR} \simeq  5.87 \times 10^{-7}  \left(\frac{2 \textrm{ keV}}{m_\textrm{WDM}} \right).
\end{equation}

The free streaming scale for WDM is given by,
\begin{equation}
\lambda_\textrm{fs} (\tau) =  \frac{v(\tau)}{a H} \simeq \frac{v(\tau)}{H_0 \Omega_r^{1/2}}a,
\end{equation}
where we can approximate $v(\tau) = c/\sqrt{3}$ when the DM is relativistic and as $v(\tau) = \sqrt{\frac{5 T }{3 m_\textrm{WDM}} }$ after the DM becomes non-relativistic. The last equality in the equation above is valid only in the radiation dominated era.

Thus, the free streaming scale at the onset of the non-relativistic era is given by
\begin{equation}
\label{eq:WDMfsscale}
\lambda_\textrm{fs} (\tau_\textrm{NR}) = 0.15 \left( \frac{a_\textrm{NR}}{5.87 \times 10^{-7}} \right) \textrm{ Mpc} =  0.15 \left(\frac{2 \textrm{ keV}}{m_\textrm{WDM}} \right) \textrm{ Mpc}.
\end{equation}

Thereafter, the free streaming scale remains constant throughout the era of radiation domination. Post MRE, the Jeans scale will fall off as $a^{-1/2}$ due to the slower growth of the Hubble length scale in the matter dominated universe.

The evolution of the Jeans scale for a 2 keV WDM candidate is shown in Fig.~\ref{fig:WDMjeanslength} along with the critical transition times. Perturbations on scales smaller than the maximum free streaming length scale will be damped relative to the standard $\Lambda$CDM scenario. Note that unlike the DR/DM scenarios that we considered in the case of the ETHOS models, we do not expect to see acoustic oscillations. However, compared to standard $\Lambda$CDM scenarios, we expect to see a damping of the power spectrum on scales smaller than the maximum free streaming length scale.

%\newpage
\bibliographystyle{utphys}
\bibliography{ethos_wdm}{}

\end{document}